\documentstyle[prb,aps,epsf]{revtex}
\begin{document}
\draft
\title{\bf  Relation between the transmission rates and the wave functions
 in the carbon nanotube junctions   }
\author{Ryo Tamura and Masaru Tsukada }
\address{Department of Physics, Graduate School of Science, University
  of Tokyo, Hongo 7-3-1, Bunkyo-ku, Tokyo 113, Japan}
\maketitle

\begin{abstract}
Electron transmission and wave functions through junctions with
 a pair of a pentagonal defect and a heptagonal defect connecting
 two metallic carbon nanotubes are analyzed
 by the analytical calculation
 with the effective mass equation.
The energy region $|E| < E_c$ is 
 considered where the channel number is kept to two.
Close relation between the transmission rate
 and the wave function is found;
the transmission rate is given by 
 the inverse squared absolute value of the wave function.
The dependence of the transmission rates on the energy
and on the size of the junction is clearly explained
 by the nature of the wave function.
Though the wave function and the transmission rate calculated 
 by the tight binding model agree well 
 with the corresponding analytical results by 
 the effective mass approximation,
 the discrepancy becomes considerable  when $|E| \simeq E_c$.
To study the origin of this discrepancy,
an efficient numerical calculation method is developed with a generalized 
transfer matrix for the tight binding model.
Their numerical results are compared with the corresponding analytical
 ones and the results show that the origin of the discrepancy comes from
the evanescent waves with the longest decay length in the tube parts.

\end{abstract}

\section{Introduction}
Recent experimental development on the carbon nanotubes, 
especially  electronic transport measurements for individual
nanotubes, \cite{experiment} has promoted much interest in the nanotubes as one-dimensional 
conductor with nanometer size. \cite{tube}
Many experimental works concern  the theoretical prediction
  that the nanotube
 becomes metallic or semiconducting according to
 its circumference. \cite{tubetheory,saito}
The conductance of the metallic nanotubes 
 with potential energy disorder has been 
 studied. \cite{impurity,nakanishiimpurity}
The junction structures with a shape of a part of a cone
 connecting two nanotubes with different radii have been also observed,
\cite{iijimajunction} and studied theoretically.\cite{tamurajunction1,tamurajunction2,saitojunction,chico}
 They are formed 
 by a pair of a pentagonal defect and a heptagonal
 defect. \cite{discl,discl2,cap}
 By composing the nanotube junctions, 
 the electronic circuits with nanometer size
 might be designed.
The atom bond network of the nanotube junctions is uniquely determined 
 by its development map with 
 the vector of the circumference of the connected thicker nanotube 
 $\vec{R}_5$  and that of the connected thinner tube $\vec{R}_7$.
We have calculated 
 the conductance with Landauer's formula, i.e., the transmission rate, 
 for the various junctions connecting two metallic nanotubes 
 by the tight binding model.\cite{tamurajunction1,tamurajunction2}
Surprisingly, the dependence of the transmission rate on the parameters,
$\vec{R}_5$, $\vec{R}_7$ and $E$, has a very simple form obeying
 the scaling law;
 in the energy region $|E| < E_c$
 where the channel number is kept to two in both the tubes,
 the transmission rate is  independent of detailed atomic arrangements
 as well as  the angle between the
 two tube axes but determined only by the two parameters;
one of the two parameters is the ratio of the circumference of the tubes $R_7/R_5$ and  the other is the scaled energy $E/E_c$.
The close relation between the transmission rate and the wave function
 in the junction part is also found by the tight binding model.\cite{tamurajunction1,tamurajunction2}
When $E=0$, the transmission rate decays with the thickness of the junction
 as $4/(2+(R_5/R_7)^3 + (R_7/R_5)^3)$ while
 the spatial decay of the corresponding wave function in the junction part 
 obeys  the same power law.\cite{tamurajunction1}
When $R_7/R_5 \simeq 0.5$, the transmission rate shows a peak
 structure as a function of the energy,
 while the corresponding wave function shows a resonant feature, i.e.,
 its amplitude is enhanced in  the middle of the junction part.\cite{tamurajunction2}

Recently,  Matsumura and Ando have confirmed the power law decay 
 for $E=0$ by using the effective mass theory.\cite{matsumura}
We generalized their discussion to more general energy region $|E|< E_c$,
 and obtained the complete analytical form of the transmission rate
 with the two parameters, $R_7/R_5$ and $|E|/E_c$. \cite{pre}
 The agreement of the analytical
 transmission rate with the numerical
 one is found to be fairly good so far as $|E|$ is not very close to $E_c$.
By our generalization of the effective mass approximation,
 the band structures of the periodic
 multiple nanotube junctions, which are called the 'helically
 coiled nanotube' in Ref.\cite{akagi} , can  be also obtained as closed
 analytical forms and their relation to the configuration of the pentagon
 and the heptagon are clearly explained based on the symmetry features.
\cite{pre2}
The effective mass theory  is derived from the tight binding model and
 is valid when the energy is near zero.
It is suitable
 to obtain the analytical results and to explain the origin of the scaling law
 because it treats not the discrete space but the continuous space, i.e.,
 it does not necessitate the detailed atomic 
 structures of the honeycomb lattice. \cite{nakanishiimpurity,matsumura,pre,pre2,ajiki,divicenzo}
Though the analytical transmission rate of the nanotube junction by the
 effective mass theory
 has been studied
 in detail in this way,
 the close relation between the transmission rate and the wave function
 observed in the numerical result is not clarified 
 by the effective mass theory yet. 
 In this paper, the spatial variations of the wave function 
 is expressed unambiguously by
 the effective mass theory and it gives an intuitive
 explanation for the dependence of the transmission
 rate on $R_7/R_5$.
 Furthermore it will be also found that its dependence  on the
 energy has close relation to the other parameter
 $|E|/E_c$ of the scaling law. These points are discussed in section III.
 The concise expression of the wave function characterized by
 only the two parameters will be useful when the STM images of
 the nanotube junction are analyzed.

On the other hand, the discrepancy between the numerical transmission
 rate with the tight binding theory
 and the analytical one by the effective mass approximation
 become considerable when $|E|$ is
 very close to $E_c$; the numerical transmission rate shows a sharp
 dip  there while the analytical one does not show.
Since the effective mass theory is an approximation derived
 from the tight binding model, the numerical result by the
 tight binding model is considered to be correct one.
To discuss the origin of the sharp dip, the wave functions
calculated by the tight binding model have to be compared with
 those calculated by the effective mass theory.
To calculate the wave function by the tight binding model
 more efficiently, 
a new numerical method  is developed in section II.
From the comparison of the wave function calculated by the
 method in section II with the analytical one,
 we found in section IV that the evanescent waves in the tube parts
 and the defect levels at the pentagon and the heptagon
 have important roles to form the dip structures.

\section{The conditioned transfer matrix method}

Fig.\ref{tenkaibars} shows the development map of the nanotube junction.
It is characterized
 by bars representing zig-zag segments of the C-C bond network
 in the circumferential direction of the single junction.
  They are aligned and numbered along the direction $\vec{e}_1-\vec{e}_2$ where $\vec{e}_1$ and $\vec{e}_2$ are the basic translation vectors of
 the graphite plane.
Each bar is connected with adjacent bars by the remaining C-C bonds. 
The network is rolled up so that the atoms denoted by $j$
 in the bottom of the $j$'th bar and the top one
 of the same bar denoted by $j'$ 
 become the identical one.
The $j$ and $j'$ sites are shown by the filled  and the open circles in Fig.\ref{tenkaibars}, respectively.
When $j \leq 0$, the $j$'th bar is formed by the $(m+n)$ unit cells
 of the 2D graphite, where 
 the $m$ unit cells are aligned along $\vec{e}_1$ and 
 the others are aligned along $\vec{e}_2$. 
 So the $(m,n)$ tube defined in Ref.\cite{saito} is formed there.
 When  $ 1 \leq j \leq l_1$, the $j$'th bar
is defined as such that made by adding  $(j-1)$ unit cells and one atoms along the direction of $\vec{e_1}$  to the bottom of the zeroth bar.
 For $ j \geq l_1+1$, $j$'th  bar  is made from the $l_1$'th  bar 
 by exchanging $(l_2-1)$ unit cells and one atoms in its bottom
 aligned along $\vec{e_1}$ 
 with $l_2$ unit cells aligned along $\vec{e_2}$.
The network made in this way 
 represents the junction of the $(m,n)$ tube $ (j \leq 0)$
and the $(m_2,n_2)$ tube $ (j \geq l_1+1)$ 
, where $ l_1=(m_2+n_2)-(m+n) \geq 0$ and $ l_2 = n_2-n \geq 0$.
This junction is called an $(m,n)$-$(m_2,n_2)$ junction hereafter.
Then Fig.\ref{tenkaibars} corresponds to the (2,2)-(2,5) junction.
A 7-membered ring is formed at the bottom of the 1st bar and a 5-membered ring
 is introduced between the $l_1$'th bar and the $(l_1+1)$'th bar.
 There are only 6 membered rings elsewhere.

The number of the bonds connecting the $(j-1)$'th bar and the $j$'th bar
is denoted by $b_j$ as
\begin{eqnarray}
b_j &= & m+n    \;\; (j \leq 0) \nonumber \\
b_j & =& m+n+j-1  \;\;( 1 \leq j \leq l_1)\nonumber \\
b_j &=& m_2+n_2  \;\; (l_1+1 \leq j )\;\;.
\end{eqnarray}
The amplitudes of the wave function in the $j$'th bar are represented by the
vector $\vec{c}_j= ^t\,(c_{j,1},c_{j,2},\cdots) $.
Its component $c_{j,i}$ can be classified to the two
 groups according to whether the corresponding site connect
 with the right $(j+1)$'th bar
 or with the left $(j-1)$'th bar.
These two groups are denoted by $r_j$ and $l_j$, respectively, 
as is shown in Fig.\ref{rlbars}.
Since each site in $j$'th bar belongs
  to only and necessarily one of the two groups,
 the dimension of $\vec{c}_j$ equals $b_j+b_{j+1}$, which are denoted
 by $d_j$ hereafter.
The bonds between the $(j-1)$'th bar and the $j$'th bar are
 numbered as Fig.\ref{rlbars} from the bottom to the top.
According to this way of numbering, 
the amplitudes are also numbered by
 $r_{j-1,i}$ and  $l_{j,i}$, which have the
 bond numbered by $i$ ($i=1,2,\cdots, b_j$).
Then the tight binding equation become
\begin{equation}
E r_{j-1,i}= \gamma (l_{j,i} +  \sum_{i'}c_{j-1,i'})
\label{tightr}
\end{equation}
and 
\begin{equation}
E l_{j,i}= \gamma (r_{j-1,i} +  \sum_{i''}c_{j,i''} ) 
\label{tightl}
\end{equation}
 where  
 the first and the other terms in the right hand sides
 represent the bonds between 
 the neighboring bars and those within the bar, respectively.
Here we use the tight binding model including only $\pi$ orbitals 
with common hopping integral, $\gamma ( \simeq -2.7$eV), 
 and common site energy chosen to be zero.
 In order to get more accurate results, the effects from the mixing with $\sigma$ orbitals caused by curvature of the graphitic plane also have to be considered. But we believe that this tight binding model
 gives a transparent view about this system
 and qualitative valid results.
Another reason why this model is used is that the purpose of
 this paper is to focus on the effects from the ${\it connectivity}$ of the
 bond network rather than those from the curvature.

The tight binding equation, eqs. (\ref{tightr}) and (\ref{tightl})
can be summarized by the matrix form
 as
\begin{equation}
 A_j\vec{c}_{j-1} +  B_j\vec{c}_{j} = 0 \;\;,
\label{ajbj}
\end{equation}
where $ A_j$ and $ B_j$ are a $2b_j \times d_j$ matrix
 and a $2b_j \times d_{j+1}$ matrix, respectively.
In the left (right) tube pars, i.e., when $j \leq 0$ ($j \geq l_1+2$), 
 these matrixes become constant matrixes,  $ A_L$ and $ B_L$
 ($ A_R$ and $ B_R$).
The transfer matrixes for the tube parts
 are obtained as $T_L=- A_L^{-1}  B_L$ for the left tube
 and $T_R=- A_R^{-1} B_R$ for the right tube.
The eigen values and eigen vectors of $T_{\mu}$ are classified into
 two groups as $\{ \beta_i^{\mu},\psi_{+i}^{\mu} \}$
 and $\{ 1/\beta_i^{\mu},\psi_{-i}^{\mu} \}$.
 Hereafter, $\mu =R$ and $\mu =L$ represent the left thinner tube
 and the right thicker tube, respectively.
The propagating waves in the former group,
 which are $ \psi_{+i}$ with $|\beta_i|=1$,
 are assigned to $i=1, \cdots, n_{\mu}$,
 where  $n_{\mu}$ is called the channel number of the
 corresponding tube.
They carry the probability flow with 
 the positive velocity, $v_i$.
The propagating waves in the latter group, $\psi_{-i}=\psi_i^*$, carry that
 with the negative velocity, $-v_i$.
From now on, the normalized extended states with unit flow,
 $\tilde{\psi}_i=\psi_i/\sqrt{v_i}$ and  $\tilde{\psi}_{-i}=\psi_{-i}/\sqrt{v_i}$ are used instead of $\psi_i$ and $\psi_{-i}$.
The other states for $ |i| > n_{\mu}$ are evanescent waves
 and the sign of $i$ means the direction along which they
 decay exponentially, i.e., $|\beta_i^{\mu}| < 1$.

The wave function in the tube parts are represented by superposition
 of these eigen vectors and its coefficients are represented
 by $\vec{x}$ as
\begin{equation}
\vec{c}_j=U_+^{\mu}(\Lambda^{\mu})^{+j}\vec{x}_+^{\mu} +
    U_-^{\mu}(\Lambda^{\mu})^{-j}\vec{x}_-^{\mu} \;\;,
\end{equation}
where the $i$'th column of $U_{\pm}^{\mu}$ is the normalized
 eigen vector $\tilde{\psi}_{\pm}^{\mu}$, and $\Lambda^{\mu}$
 is the diagonal matrix whose $(i,i)$ element is the eigen
 value $\beta_i^{\mu}$.

Unlike the tube parts, the 'usual' square transfer matrix cannot be obtained
 in the junction part,
 because the dimension of $\vec{c}_j$ and that of $\vec{c}_{j+1}$
 are different, i.e., the $A_j$ for $1 \leq j \leq l_1$
 becomes a rectangle matrix which has no inverse matrix.
 Nevertheless there is  a $d_{j-1} \times  2b_j$ matrix 
 which is similar 
 to the inverse matrix of $A_j$. It is denoted by $\tilde{A}_j$ and
 defined by
\begin{equation}
\tilde{A}_j \equiv (A_j^{\dag}A_j)^{-1}A_j^{\dag}\;\;.
\label{tildeA}
\end{equation}
This matrix satisfies the condition $\tilde{A}_jA_j=1$
 but $A_j\tilde{A}_j \neq 1$.
By this pseudo-inverse matrix, we can define the
 'transfer matrix', $T_j \equiv -\tilde{A}_jB_j$.
If eq.(\ref{ajbj}) is satisfied, then
\begin{equation}
\vec{c}_{j-1}= T_j \vec{c}_{j}
\label{Tj}
\end{equation}
is satisfied, but its converse does not hold generally.
It means that $\vec{c}_{j+1}$ in eq.(\ref{Tj})
 cannot be chosen to be arbitrary
 but has to satisfy 
\begin{equation}
(B_j + A_jT_j)\vec{c}_{j}=0\;\;.
\label{condition}
\end{equation}
To make eq.(\ref{Tj}) equivalent to eq.(\ref{ajbj}),
the condition (\ref{condition}) is necessary,
 so we call $T_j$  a 'conditioned transfer matrix' hereafter.
In our problem, the number of the independent rows of $(B_j + A_jT_j)$
 is only one, so eq.(\ref{condition}) can be written as
\begin{equation}
^t\,\vec{s}_j \cdot \vec{c}_{j}= 0\;\;,
\label{condition2}
\end{equation}
where $^t\,\vec{s}_j$ is one of the nonzero rows of $(B_j + A_jT_j)$.
Multiplying $T_j$ generates the coefficient at the bar with the decreased $j$,
 so we define the positive direction along which $j$ decreases,
 i.e., from the right thicker tube to the left thinner tube.

By using the inverse matrix $^t\,(^t\,V_+^L,^t\,V_-^L) \equiv (U_+^L, U_-^L)^{-1}$ and $K_j \equiv T_j T_{j+1} \cdots T_{l_1}$,
 the relation between $\vec{c}_{l_1+1}$ and the right going waves 
in the thinner tube $\vec{x}_-^L$ are summarized as
\begin{equation}
\left( \begin{array}{c}
V^L_-K_0 \\
^t\,\vec{s}_{\l_1+1} \\
^t\,\vec{s}_{l_1}K_{l_1} \\
^t\,\vec{s}_{l_1-1}K_{l_1-1} \\
\vdots \\
^t\,\vec{s}_{2}K_{2} \\
\end{array} \right)\vec{c}_{l_1+1}=
\left( \begin{array}{c}
\vec{x}_-^{L} \\
0 \\
0 \\
0 \\
\vdots \\
0 \\
\end{array} \right)\;\;.
\label{Q}
\end{equation}
The $(m_2+ n_2) \times (2m_2+2n_2)$
 matrix in the left hand side of eq.(\ref{Q})
 is represented by $Q$ hereafter.

Now we define
all the necessary things to calculate the transmission rate
 and the wave function.
When the electron is incident from the right and transmitted into
 the thinner tube
, i.e., when $\vec{x}_+^R \neq 0$ and $\vec{x}_-^L = 0$,
 the reflected wave $\vec{x}^R_-$ and the transmitted wave $\vec{x}^L_+$
 are obtained from
\begin{equation}
\vec{x}^R_- = -(QU_-^R)^{-1}(QU_+^R)\vec{x}^R_+
\label{reflect+}
\end{equation}
and
\begin{equation}
\vec{x}^L_+ = V^L_+K_0\left( U_+^R -U_-^R(QU_-^R)^{-1}(QU_+^R) \right)\vec{x}^R_+\;\;,
\label{transmission+}
\end{equation}
respectively.
The corresponding wave function in the junction part,
 i.e., $\vec{c}_j$ for $1 \leq j \leq l_1$ is given by
\begin{equation}
\vec{c}_j = K_j \left( U_+^R -U_-^R(QU_-^R)^{-1}(QU_+^R) \right)\vec{x}^R_+\;\;.
\label{wavefunction+}
\end{equation}

For the inverse direction of the incident electron,
 $\vec{x}_+^R = 0$ and $\vec{x}_-^L \neq 0$,
 the reflected wave $\vec{x}_+^L$
and the transmitted wave $\vec{x}_-^R$ 
 are obtained 
from
\begin{equation}
\vec{x}^L_+ = V^L_+K_0 U_-^R(QU_-^R)^{-1}
\left( \begin{array}{c}
\vec{x}_-^{L} \\
0 \\
\end{array} \right)\;\;,
\label{reflect-}
\end{equation}
and
\begin{equation}
\vec{x}^R_- = (QU_-^R)^{-1}
\left( \begin{array}{c}
\vec{x}_-^{L} \\
0 \\
\end{array} \right)\;\;.
\label{transmission-}
\end{equation}
In this case, the wave function in the junction part is given by
\begin{equation}
\vec{c}_j = K_j U_-^R(QU_-^R)^{-1}
\left( \begin{array}{c}
\vec{x}_-^{L} \\
0 \\
\end{array} \right)\;\;.
\label{wavefunction-}
\end{equation}
The transmission rate is 
obtained from eq.(\ref{transmission+}) or eq.(\ref{transmission-}),
 while the reflection rate is obtained from eq.(\ref{reflect+})
 or eq.(\ref{reflect-}) from the matrix elements corresponding
to the open channel, i.e., $x^{\mu}_{\pm i}$ with $i=1 \sim n_{\mu}$.
We have confirmed that unitarity  holds very well
 in the conditioned transfer matrix method.
Agreement of their results with
 those  calculated by usual recursive Green's function methods
is also quite good.
In both the methods, matrix inversions are necessary about $l_1$
 times, but they can be done much faster in the conditioned
 transfer matrix method than in 
 the recursive Green's function method,
 because matrixes which have to be inverted
 are real sparse symmetric matrixes, $A_j^{\dag}A_j$, in the former
 method while those in the latter method are complex matrixes
 with no symmetry.
Therefore the 'conditioned transfer matrix'
 method is much faster than 
the recursive Green's function method.
 This efficiency is expected to be more important
 when more general related problems are calculated,
 e.g., those including electron-phonon interaction or
 electron-electron interaction self-consistently.
Another advantage of the conditioned transfer matrix
 method over the recursive Green's function method,
 which is not essential but practically important,
 is that it is  more intuitive and easy to be implemented.
%The conditioned transfer matrix method
% has also  a disadvantageous point that
% its possible number of the step $l_1$
% is usually less than that of the recursive Green's function
% method.
%But this difference is not so large.
%For example, the maximum value of $i$ of the (14,5)-$(7+i,13+i)$
% junctions to maintain accuracy of three columns is
% $i=9$ in the former method and $i=15$ by the latter method,
%in our program.

In section IV, the wave functions are calculated numerically
 by the conditioned transfer matrix method
 and compared to those calculated analytically by the
 effective mass equation.

\section{Effective mass theory and its application to the single nanotube
 junction}
Fig. \ref{junctiontenkai} shows a 
 development map of the nanotube junction neglecting the atomic sites.
The vectors  $\vec{R}_5 $ and $\vec{R}_7$
 represent the circumferences of the thicker tube and the thinner 
 tube, respectively.
Lines $E_7P_7P_5E_5$ and lines $F_7Q_7Q_5F_5$ are sticked each other
 so that the points connected by $\vec{R}_j$ become identical ($j=5, 7$). 
Then $P_5(=Q_5)$ and $P_7(=Q_7)$ turn out to be
 the centers of  a heptagonal defect 
 and a pentagonal defect, respectively.
These points are marked also in Fig.\ref{tenkaibars}.
 Thus from now on, the indices '7' and '5' are
 used to represent the thinner and the thicker tube, respectively.

The equilateral triangles '$\Delta  OP_7Q_7$'
 and '$\Delta OP_5Q_5$'
 with bases '$P_7Q_7$' and '$P_5Q_5$'
 have a common apex $ O$,
 which is chosen to be the origin of the coordinate
$(x,y)$ in this paper. \cite{saitojunction}
The origin $O$ lies between $-(m+n)$'th bar and $-(m+n-1)$'th bar
 , since $\vec{Q_7 O}= m \vec{e}_2 + n( \vec{e}_2-\vec{e}_1)$.

Then the position of a general site, $\vec{r}$,
 can be labeled by $(s,q_1,q_2)$,  where $q_1$ and
 $q_2$ are integer components representing the position of the
 unit cell, $\vec{q}\equiv q_1\vec{e}_1+q_2\vec{e}_2$, and $s$ represents the sublattice,
 $s=A, B$.
Relation between the position vector of the site, $\vec{r}$,
 and its label $(s,q_1,q_2)$ is written by
 \begin{eqnarray}
 \vec{r}(A,q_1,q_2)& =& \vec{q}+(\vec{e}_1+\vec{e}_2)/3 \nonumber\\
 & = & \frac{\sqrt{3}}{2}(q_1+q_2+\frac{2}{3})\vec{e}_x +\frac{1}{2}(q_2-q_1)\vec{e}_y 
\label{labelA}
\end{eqnarray}
 and 
 \begin{eqnarray}
 \vec{r}(B,q_1,q_2)& =& \vec{q}+2(\vec{e}_1+\vec{e}_2)/3 \nonumber\\
 & = & \frac{\sqrt{3}}{2}(q_1+q_2+\frac{4}{3})\vec{e}_x +\frac{1}{2}(q_2-q_1)\vec{e}_y \;.
\label{labelB}
\end{eqnarray}
In the above $\vec{e}_{x}\equiv(\vec{e}_1+\vec{e}_2)/\sqrt{3} $ and  $\vec{e}_{y}\equiv \vec{e}_2-\vec{e}_1 $, which define the coordinate
 system $(x,y)$  shown in Fig.\ref{tenkaibars}.
The amplitude of the wave function at these sites is denoted by
 $\psi(s,q_1,q_2)$.
In the effective mass theory, the wave function is represented  by
\begin{equation}
\psi(s, q_1,q_2)=F_s^{K}(\vec{r})w^{(q_1-q_2)}
 +F_s^{K'}(\vec{r})w^{(q_2-q_1)}\;\;\;(s=A,B).
\label{defFA}
\end{equation}
Here $w \equiv \exp( i2\pi/3)$ and $\vec{r}$ is defined by eqs.(\ref{labelA}) and (\ref{labelB}) as a function of the label $(s,q_1,q_2)$.
In eq.(\ref{defFA}),
 $F_{A,B}^{K,K'}$, $w^{(q_1-q_2)}$ and $w^{(q_2-q_1)}$  are the
 envelop wave functions, 
the wave function of the Bloch states at the $K$ and
 the $K'$ point , respectively.
This wave function is expressed by a vector, $ \vec{\psi} =
 (F_A^K(\vec{r}),F_B^K(\vec{r}+\vec{e}_x/\sqrt{3}),F_A^{K'}(\vec{r}),F_B^{K'}(\vec{r}+\vec{e}_x/\sqrt{3}))$, hereafter.
This definition of F's is different from
 other references \cite{nakanishiimpurity,pre,ajiki,matsumura} by certain factors.
The reason why this definition is used is that the  representation
 of the time reversal operation $I$,  $I\psi = \psi^*$,
  becomes simpler as
\begin{equation}
I(F_A^K,F_B^K,F_A^{K'},F_B^{K'})=((F_A^{K'})^*,(F_B^{K'})^*,(F_A^K)^*,(F_B^K)^*)\;\;.
\label{timerev}
\end{equation}
By using eq.(\ref{defFA}), the effective mass equation are derived 
from the tight binding equation used in the preceding
 section.
When energy $E$ is zero,
  $w^{q_1-q_2}$ and $w^{q_2-q_1}$ are solutions of the tight 
 binding model, so that the corresponding solutions of 
 the envelop functions $F$ are constant.
Thus when $E$ is not zero but close to zero,
spatial variation of the envelop functions is slow compared to the lattice constant, $|\vec{e}_x| = |\vec{e}_y| \equiv a \simeq 0.25$nm.
In this case, it is a good approximation to take only the first order term in
 the Taylor expansion of the envelop function as 
$F(\vec{r}+\vec{e}_x) \simeq (1 +a \partial_{x}) F(\vec{r})$.
 From this approximation, one obtains \cite{nakanishiimpurity,ajiki,divicenzo}
\begin{equation}
(-i\partial_{y}+ \partial_{x})F^K_B(\vec{r}+\vec{b})=  
\epsilon F^K_A(\vec{r})
\label{kpKB}
\end{equation}
\begin{equation}
(-i\partial_{y}- \partial_{x})F^K_A(\vec{r})= \epsilon F^K_B(\vec{r}+\vec{b})\;\;.
\label{kpKA}
\end{equation}
\begin{equation}
(i\partial_{y}+ \partial_{x})F^{K'}_B(\vec{r}+\vec{b})= \epsilon 
F^{K'}_A(\vec{r})
\label{kpK'B}
\end{equation}
\begin{equation}
(i\partial_{y}- \partial_{x})F^{K'}_A(\vec{r})= \epsilon F^{K'}_B(\vec{r}+\vec{b})\;\;,
\label{kpK'A}
\end{equation}
where $\epsilon = 2 E/(\sqrt{3} \gamma a)$ and 
 $\vec{b} \equiv \vec{e}_x/\sqrt{3}$.
When the plane wave solution $F_A^K, F_B^K \propto 
\exp(\pm i\vec{k}\cdot\vec{r})$ is used in eqs.(\ref{kpKB}) and 
 (\ref{kpKA}), one can get  linear isotropic dispersion relation,
\begin{equation}
k \equiv |\vec{k}| = |\epsilon|\;\;,
\label{dispersion}
\end{equation}
 and $F_B^K(\vec{r}+\vec{b})/F_A^K(\vec{r})= \pm \exp(i\eta)$,
 where $(k_x,k_y)=(-\epsilon \sin \eta,\epsilon \cos \eta)$, i.e.,
 $\eta$ is the angle of $\vec{k}$ with respect 
 to the $y$ axis measured anti-clockwise.
It follows that the corresponding wave function, $\vec{\psi}_{K\pm}$, is
 written as
\begin{equation}
\vec{\psi}_{K\pm}=(e^{-i\eta/2}, 
\pm e^{i\eta/2},0,0 ) e^{\pm i(\vec{k} \cdot \vec{r})}
\;\;.
\label{propK}
\end{equation}
The wave number must satisfy the boundary condition in the metallic
 nanotube parts as 
\begin{equation}
\vec{k}\cdot\vec{R_j}= 2 \pi l_j \;\;(j=5,7)\;,
\label{tubebound}
\end{equation}
 where $l_j$ is an integer  representing the number of oscillations of the envelop function around the circumferences.
As for the semiconducting tubes, the boundary condition becomes
different  from eq.(\ref{tubebound}) \cite{ajiki}, 
but we concentrate our discussion
 to the metallic nanotubes in this paper.
 The upper sign and the lower sign in eq.(\ref{propK})
 represent the direction of the propagating waves.
When $E$ is close to zero, i.e., $k \simeq 0$, only possible number
 of $l_j$ is zero. It means that $\vec{k}$ is perpendicular to $\vec{R}_j$ so that  $\eta$ is the angle of $\vec{R}_j$ with respect to $x$ axis.
Within the effective mass theory scheme,
the possible maximum value of $l_5$ ($l_7$)
 is the channel number of the thicker tube (the thinner tube).
Therefore the range of the energy where the channel number is kept to two
 in the thicker tube  ( the thinner tube) 
 is $|E/\gamma| < \sqrt{3} \pi a /R_5$ ($|E/\gamma| < \sqrt{3} \pi  a /R_7$). Discussions in this paper are concentrated to  the  energy region
 where the channel number is kept to two in both the tubes.
It is represented by $|E| < (\sqrt{3}\pi|\gamma|)(a/R_5) \equiv E_c$ and $E_c$ is called a threshold energy hereafter.
 From the propagating waves near the $K$ point, $\vec{\psi}_{K\pm}$,
 the other propagating waves $\psi_{K'\pm}$ 
 are obtained by the time reversal operation (\ref{timerev})
 as
\begin{equation}
\vec{\psi}_{K'\pm}=I \vec{\psi}_{K\mp}=(0,0,e^{i\eta/2}, 
\mp e^{-i\eta/2}) e^{\pm i(\vec{k} \cdot \vec{r})}
\;\;.
\label{propK'}
\end{equation}
Note that the direction of the propagation is reversed by the time reversal
 operation $I$.
In order to discuss the wave function in the junction
part, the polar coordinate $(r,\theta)$ is useful.
Its relation to the coordinate $(x,y)$ is the 
usual one, i.e., $ r=\sqrt{x^2+y^2}$, $\tan \theta =y/x$.
Then the wave function satisfies the wave equation 
$r^2(\partial_x^2+\partial_y^2+k^2)F=(z^2\partial_z^2+z\partial_z+\partial_{\theta}^2+z^2) F = 0$, where $z=kr$.
The solution is represented by Bessel functions $J_m$ and Neumann functions
 $N_m$  as 
\begin{equation}
F= \sum_{m=-\infty}^{\infty} e^{im\theta}(c_m J_{|m|}(z)+d_m N_{|m|}(z)) \;\;.
\label{JmNm}
\end{equation}

In Fig.\ref{tenkaibars}, site $i$ is identical with site $i'$ 
for $i=1,\cdots l_1$, while corresponding labels $(s,q_1,q_2)$
are different between $i$ and $i'$; 
that for $i$ is $(B,1,-4-i)$ and that for $i'$ is $(A,3+i,-2-i)$.
In this way, the wave function in the junction part must satisfy
 the condition
\begin{eqnarray}
\psi(A,q_1,q_2)= \psi(B,q_1+q_2,-q_1-1)\;\;.
\label{junctbound}
\end{eqnarray}
Here transformation of the label $(A,q_1,q_2) \rightarrow
 (B,q_1+q_2+1,-q_1-1)$ is equivalent to clock-wise rotation by $\pi/3$
 with respect to the origin $O$.
From eqs. (\ref{junctbound}) and  (\ref{defFA}),
the boundary conditions in the junction part are derived as
\begin{equation}
F^{K'}_A(r,\theta+\pi/3)=wF^K_B(r,\theta)\;\;,
\label{bound1}
\end{equation}
and
\begin{equation}
F^{K}_A(r,\theta+\pi/3)=\frac{1}{w}F^{K'}_B(r,\theta)\;\;.
\label{bound2}
\end{equation}
The other boundary conditions
\begin{equation}
F^{K}_B(r,\theta+\pi/3)=wF^{K'}_A(r,\theta)\;\;,
\label{bound3}
\end{equation}
and
\begin{equation}
F^{K'}_B(r,\theta+\pi/3)=\frac{1}{w}F^{K}_A(r,\theta)\;\;.
\label{bound4}
\end{equation}
are also obtained in the same way.
The same boundary conditions are firstly discussed by
 Matsumura and Ando.\cite{matsumura}
 Difference of eqs.  (\ref{bound1}) -  (\ref{bound4})
  from those of Matsumura and
 Ando by certain factors is due to the difference of the definition of 
 $F_{A,B}^{K,K'}$.
From eqs. (\ref{bound1}) and  (\ref{bound2}), terms in eq. (\ref{JmNm})
for $F^{K'}_A$ and $F^K_B$ are not zero only when $m=3 p+2 \;\;(p=$integer
).
 Because the open channel $l_j=0$ in the tube parts
 is spatially uniform along the circumference,
 it is better fitted to the components with smaller $|m|$ in eq. (\ref{JmNm}) 
 than to those with larger $|m|$.
So we assume that one can neglect all the terms except those with $p=0$ and $p=-1$ in eq. (\ref{JmNm}) (Assumption I). Then the wave functions can be written as
\begin{equation}
F^{K'}_A=e^{2i\theta} f_2(z)+e^{-i\theta} f_1(z),
\label{fk'a}
\end{equation}

and

\begin{equation}
F^{K}_B=e^{2i\theta }f_2(z)-e^{-i\theta} f_1(z)\;\;,
\label{fkb}
\end{equation}

where

\begin{equation}
f_m(z)=c_mJ_m(z)+d_mN_m(z) \;\;(m=1,2).
\label{fm}
\end{equation}

From eqs. (\ref{kpKB}) and  (\ref{kpK'A}), the other two wave functions $F^{K'}_B$ and $F^K_A$ can be derived from $F^{K'}_A$ and $F^K_B$ as

\begin{equation}
F^{K'}_B= \frac{\epsilon}{|\epsilon|}\left(
 -e^{i\theta} \tilde{f}_2(z)+e^{-i2\theta} \tilde{f}_1(z) \right) \;\;,
\label{fk'b}
\end{equation}

\begin{equation}
F^{K}_A= \frac{\epsilon}{|\epsilon|}
\left( e^{i\theta} \tilde{f}_2(z)+e
^{-i2\theta} \tilde{f}_1(z) \right)\;\;,
\label{fka}
\end{equation}
where
\begin{eqnarray}
\tilde{f}_1(z)=c_1J_2(z)+d_1N_2(z),\nonumber \\
\tilde{f}_2(z)=c_2J_1(z)+d_2N_1(z)\;\;.
\label{tildef}
\end{eqnarray}
In the above we used the recursion formula of the Bessel functions and Neumann
 functions.\cite{noteepsilon}
It is easily confirmed that eqs. (\ref{fk'b}) and (\ref{fka}) satisfy
 the boundary conditions eqs. (\ref{bound3}) and (\ref{bound4}).
 The amplitude of the open channel in the tube, which is denoted by $\alpha$,  
 is obtained from eq.(\ref{propK}) as
\begin{equation}
\alpha_{j\pm}^K=\frac{1}{2\sqrt{R_j}}\int_{Q_j}^{P_j} dx^{(j)} 
(e^{i\frac{\eta_j}{2}}F^K_A \pm  e^{-i\frac{\eta_j}{2}}F_B^K)\;\;(j=5,7)\;\;,
\label{alpha}
\end{equation}
for the $K$ point.
The value of $|\alpha|^2$ equals the probability flow as shown in Appendix.
The indices $+$ and $-$ mean the direction along which the electron waves
propagate.
The path of the integral of eq.(\ref{alpha}) is the straight
 line $P_jQ_j$, the angle of which with respect to $x$ axis 
 is denoted by $\eta_j$.
Equations for $\alpha^{K'}_{j\pm}$  are obtained from eq. (\ref{alpha}) by replacing $\pm$, $\eta_j$  and $K$ in the r. h. s with $\mp$, $-\eta_j$ and $K'$, respectively.
To simplify the calculation, the integrations
in the above  are transformed as
\begin{equation}\int_{Q_j}^{P_j} dx^{(j)} \rightarrow R_j \int_{-\frac{2}{3}\pi+\eta_j}
^ {-\frac{\pi}{3}+\eta_j}  d\theta\;\;.
\end{equation}
If variation of the wave function  along the radial directions
 is slow near $r=R_j$, this replacement can be allowed (Assumption II).
The relation between the amplitudes of the open channel in each tube,
$\vec{\alpha}_j=\; ^t\,(\alpha_{j+}^K,\alpha_{j+}^{K'},\alpha_{j-}^{K'},
\alpha_{j-}^{K} )$, and the coefficients representing
 the wave functions in the junction part,
 $\vec{g}=\; ^t\,(c_2,d_2,c_1,d_1 )$, are summarized in the followings.

\begin{equation}
\vec{\alpha}_j=\frac{1}{2}\sqrt{R_j}P(\eta_j) M \Lambda(\eta_j) L(k R_j ) \vec{g} \equiv Y_j \vec{g}\;\;,
\label{a5mat}
\end{equation}
where $M$ is a constant matrix given by \cite{noteepsilon}
\begin{equation}
M=
\left( \begin{array}{cccc}
-i\frac{\epsilon}{|\epsilon|}& 0 & 0 & -\frac{\sqrt{3}}{2}\frac{\epsilon}{|\epsilon|}\\
0 & -\frac{\sqrt{3}}{2} & -i& 0 \\
0 &  -\frac{\sqrt{3}}{2}& i & 0 \\
i\frac{\epsilon}{|\epsilon|} & 0 & 0 &  -\frac{\sqrt{3}}{2}\frac{\epsilon}{|\epsilon|}
\end{array} \right)\;\;.
\label{M}
\end{equation}
$\Lambda(\eta)$ is a diagonal matrix, where $\Lambda_{1,1}=\Lambda_{3,3}^*=e^{i \eta}$ and $\Lambda_{2,2}=\Lambda_{4,4}^*=e^{2i \eta}$.
$P(\eta)$ is defined by eq. (\ref{alpha}) as 
\begin{equation}
P(\eta) =
\left( \begin{array}{cccc}
e^{i\frac{\eta}{2}},& e^{-i\frac{\eta}{2}}, &0, &0\\
0, & 0, &e^{-i\frac{\eta}{2}},& -e^{i\frac{\eta}{2}} \\
 0, & 0,& e^{-i\frac{\eta}{2}},& e^{i\frac{\eta}{2}} \\
e^{i\frac{\eta}{2}},& -e^{-i\frac{\eta}{2}}, &0, &0
\end{array} \right)\;\;.
\label{P}
\end{equation}
 The matrix elements of  $L(z)$ are $L_{11}=L_{33}=J_1(z)$, $L_{12}=L_{34}=
N_1(z)$, $L_{21}=L_{43}=J_2(z)$ and $L_{22}=L_{44}=N_2(z)$. The other matrix
 elements of $L(z)$ are zero.
From eq. (\ref{a5mat}), the relation 
  between $\vec{\alpha}_7$ and $\vec{\alpha}_5$
 is given by $\vec{\alpha}_7=Y_7Y_5^{-1}\vec{\alpha}_5$ with
 the  three parameters,  $kR_7$, $kR_5$ and $\phi \equiv \eta_7-\eta_5$ 
 (angle between $\vec{R}_5$ and $\vec{R}_7$ in the development map).
This reads   as
\begin{equation}
\left( \begin{array}{c}
\vec{\alpha}_{7+} \\
\vec{\alpha}_{7-}\\
\end{array} \right)
 =
\left( \begin{array}{cc}
t_{1} ,& t_{2}^* \\
t_{2} ,& t_{1}^* \\
\end{array} \right)
\left( \begin{array}{c}
\vec{\alpha}_{5+} \\
\vec{\alpha}_{5-}\\
\end{array} \right)\;\;.
\label{transfer}
\end{equation}
where
\begin{equation}
t_1 =h_+
\left( \begin{array}{cc}
\cos(\frac{3}{2}\phi),& i \sin(\frac{3}{2}\phi) \\
i\sin(\frac{3}{2}\phi),& \cos(\frac{3}{2}\phi) \\
\end{array} \right)\;\;,
\label{t1}
\end{equation}

\begin{equation}
t_2 =h_-
\left( \begin{array}{cc}
-\cos(\frac{3}{2}\phi),& -i \sin(\frac{3}{2}\phi)\\
i\sin(\frac{3}{2}\phi),& \cos(\frac{3}{2}\phi) \\
\end{array} \right)\;\;.
\label{t2}
\end{equation}
The factors $h_+$ and $h_-$ in eqs.(\ref{t1}) and (\ref{t2})
 are represented by
\begin{equation}
h_{\pm}= -\frac{1}{4}\left(X_{12}(kR_7,kR_5)
 \mp X_{21}(kR_7,kR_5) \right) 
 +\frac{i}{2\sqrt{3}} \left (X_{11}(kR_7,kR_5) \pm \frac{3}{4}X_{22}(kR_7,kR_5) \right)
\label{hpm}
\end{equation}
where
\begin{equation}
X_{i,j}(z_1,z_2)\equiv \sqrt{z_1z_2}\pi \{J_i (z_1) N_j(z_2)-N_i(z_1) J_j(z_2)\}\;\;.
\label{X}
\end{equation}
The considered energy region $|E|<E_c$ corresponds to the region,
 $0 < kR_5 < 2\pi$.
It can be easily confirmed analytically that 
 eq. (\ref{transfer}) satisfies the time reversal symmetry
 and unitarity.
 The transmission rate per channel denoted by $T$ is calculated from eq.(\ref{transfer})
 as 
\begin{equation}
T=1/|h_+|^2=\frac{2}{
\{(1/6)\sum_{i=1}^2\sum_{j=1}^2 
(3/4)^{i+j-2}X^2_{i,j}(kR_5,kR_7)
 \} + 1 }\;\;,
\label{sigma}
\end{equation}
 and it gives the conductance $\sigma$ as $\sigma=2T$ by Landauer's formula.
In Fig. \ref{T13-13},
 the solid lines   represent
 the values of $T$ calculated by the tight binding model,
 and the 
 dotted line and the dashed line represent those
 calculated by eq. (\ref{sigma}).
Agreement between the two methods is fairly good.\cite{pre}
When $E=0$,  $T$ becomes
\begin{equation}
T=\frac{4}{(R_5/R_7)^3 + (R_7/R_5)^3+2}\;\;.
\label{zerosigma}
\end{equation}
Eq. (\ref{zerosigma}) reproduces well  the numerical results
 in Reference.\cite{tamurajunction1}.

 We are now ready to consider the analytical calculation
 of the wave functions by the effective mass theory
  and its relation to the transmission rate.
 Let us  consider the case 
 where the two  waves $\vec{\alpha}^{(i)}_{5+} \;\; (i=1,2)$
 are incident from the thicker tube,  and there is no incidence
 from the thinner tube, i.e., $\vec{\alpha}_{7-}=0$.
As in section II, it is called the positive  incidence direction.
Whenever the two incident waves are 'orthogonal' and have the 
 same flow, \cite{note} 
 i.e., $|\vec{\alpha}_{5+}^{(1)}|=|\vec{\alpha}_{5+}^{(2)}|$
 and $^t\!\vec{\alpha}_{5+}^{(1)*}\cdot\vec{\alpha}_{5+}^{(2)}=0$,
 the corresponding two transmitted waves are also
 "orthogonal" and have  the same flow,
 because $t_1^{\dag}t_1 =  1/T$.
 Note that it does not hold generally and characteristic
 of the nanotube junction. \cite{note2,brandbyge} 

The two orthogonal transmitted waves are represented by
$\vec{\alpha}^{(1)}_{7+}=^t\!(a_1,a_2)$ and
$\vec{\alpha}^{(2)}_{7+}=^t\!(-a^*_2,a^*_1)$.
In the analytical calculation, we use different normalization from that of
 section II; we take the transmitted waves with unit flow represented by
 $|\vec{\alpha}^{(i)}_{7+}|=1$ in this section
 while the incident waves have unit
 flows, $|\vec{\alpha}^{(i)}_{5+}|=1$, in section II.
Corresponding wave functions denoted by $\psi^{(i)}$ are obtained from
$\vec{g}=Y_7^{-1}\vec{\alpha}^{(i)}_7$ and eqs.(\ref{fk'a})-(\ref{tildef}).
 They depend on the phases of $a_1$ and $a_2$,
 but the 
 sum of the squared absolute values of them, $\Psi(s,q_1,q_2) \equiv 
|\psi^{(1)}(s,q_1,q_2)|^2 + |\psi^{(2)}(s,q_1,q_2)|^2 $,
 does not depend on them and can be determined uniquely.
 Therefore we discuss $\Psi= |\psi^{(1)}|^2+|\psi^{(2)}|^2$
 rather than $|\psi^{(1)}|^2 $ and $|\psi^{(2)}|^2 $,
 hereafter.
Similar discussions are also possible  for the negative incidence direction.
The $\Psi(s,q_1,q_2)$ in the junction part is derived as
\begin{equation}
 \Psi(A,q_1,q_2)=\frac{1}{3r}
 \sum_{i=1}^2 (3/4)^{i-1}\left(X_{i,1}^2
+ X_{i,2}^2
+ 2\frac{\epsilon}{|\epsilon|}
\cos \left( \theta + \frac{2\pi}{3}(q_1-q_2) \right)X_{i,1}X_{i,2}  \right) \;\;,
 \label{psiA}
\end{equation}
and
\begin{equation}
 \Psi(B,q_1,q_2)=\frac{1}{3r}
 \sum_{i=1}^2 (3/4)^{i-1}\left(X_{i,1}^2 +X_{i,2}^2
- 2\frac{\epsilon}{|\epsilon|}
\cos \left( \theta - \frac{2\pi}{3}(q_1-q_2) \right)X_{i,1}X_{i,2}  \right) \;\;,
 \label{psiB}
\end{equation}
 where $X_{i,j}=X_{i,j}(kR_t,kr)$ is defined by eq.(\ref{X}),
 with the circumference $R_t$ of the tube into which the electron
 is transmitted.
That is to say, $R_t=R_7$ and $ kR_t < kr < kR_5 < 2\pi$ for the positive
 incidence direction,
 while  $R_t=R_5$ and $ kR_7 < kr < kR_t < 2\pi$ for the negative incidence
 direction.
Note that the squared wave function $\Psi$ in eqs.(\ref{psiA}) and 
 (\ref{psiB})
 does depend on the sign of
energy unlike the conductance in eq.(\ref{sigma}).
Note also that the sign of $E$ is inverse of that of 
 $\epsilon$,  $E/|E|= -\epsilon/|\epsilon| = \pm1$, because 
 the transfer integral $\gamma$ is negative.
In the right hand sides of eqs.(\ref{psiA}) and (\ref{psiB}),
only the third term depends on the angle $\theta$,
and it becomes zero as $E$ approaches zero, because $X_{i,i}\rightarrow 0 $.
Therefore the spatial oscillation  of $\Psi$ 
 along the $\theta$ direction
decreases with decreasing $|E|$.

There are several possible choice of the basic translation vectors,
 $\vec{e}_1$ and  $\vec{e}_2$. The wave function represented by
 eqs.(\ref{psiA}) and (\ref{psiB}) should be
invariant under the change of the choice.
When $\vec{e'}_1=\vec{e}_2$ and $\vec{e'}_2=\vec{e}_2-\vec{e}_1$ are
 used instead of  $(\vec{e}_1, \vec{e}_2)$, the corresponding
 $x'$, $y'$ axes are rotated ones by $\pi/3$ with respect to original $x$,$y$ axes as shown in Fig.\ref{tenkaibars}.
Correspondingly, the polar coordinates and the labels are transformed
 as
$(r,\theta) \rightarrow (r,\theta-\pi/3)$, $(A, q_1,q_2) \rightarrow
 (B,q_1+q_2,-q_1-1)$ and  
$(B, q_1,q_2) \rightarrow (A, q_1+q_2+1,-q_1-1)$.
For example, the site indicated by the square symbol
   near the origin $O$ in Fig.\ref{tenkaibars}
 has the two different labels, which are $(A,0,0)$  defined
  by $(\vec{e}_1,\vec{e}_2)$ and $(B,0,-1)$ 
 defined by $(\vec{e'}_1,\vec{e'}_2)$.
Under this transformation of $(r,\theta)$ and $(s,q_1,q_2)$, 
the values of eqs.(\ref{psiA}) and (\ref{psiB})
 are invariant, that is to say,
 the results are independent of the way of the labeling, or choice of
 the vectors, $\vec{e}_1$ and $\vec{e}_2$.

In order to relate the wave function to the transmission rate,
the integral of the squared wave function defined below should be
 introduced:
\begin{equation}
\Phi(kR_t,kr) \equiv \frac{9r}{14\pi}\int_Q^P d \theta  \left( \Psi(A;r, \theta) + \Psi(B;r, \theta) \right) 
 = \frac{1}{7}\sum_{i=1}^2\sum_{j=1}^2 (3/4)^{i-1}X^2_{i,j}(kR_t,kr)\;\;.
\label{norm}
\end{equation}
The third terms in r. h. s of eqs.(\ref{psiA}) and (\ref{psiB}) 
do not contribute to the integral because $q_1-q_2$ varies rapidly
as a function of $\theta$.
We call $\Phi$ defined by eq.(\ref{norm}) a radial norm, hereafter.
The radial norm is normalized so that it equals unity
 at $r=R_t$, i.e., at the exit of the transmitted wave from the junction part.
It corresponds to the unit flow of the transmitted wave.
Thus at the entrance of the incident wave into the junction part,
 the radial norm  $\Phi$
 becomes the sum of the amplitude of incident wave $1/T$ and that
 of the reflected wave $R/T=(1-T)/T$.
Accordingly, the transmission rate $T$ can be given
 by $T=2/(\Phi +1)$, where $\Phi = \Phi(kR_7,kR_5)$ for the
 positive incidence direction and $\Phi=\Phi(kR_5,kR_7)$ for the
 negative incidence direction.
Figure \ref{scheme}  is the schematic development maps
 representing this relation between
 the radial norm and the transmission rate.
It shows two junctions
 with a common thicker tube where
 the incidence direction is negative.
Though they have different transmission rates, $T_1$ and $T_2$,
their wave functions in the region CC'D'D and
 the transmitted wave are common.
From now on, we call $2/(\Phi+1)$ an 'inverse' of the radial norm
 for simplicity.
The inverse of the radial norm, $2/(\Phi(kR_7,kR_5)+1)$ ,$2/(\Phi(kR_5,kR_7)+1)$  and  the transmission rate $T$ of eq.(\ref{sigma}) are shown in Fig.\ref{transnorm} as a function of  $E/E_c$ by the dashed lines, the dotted lines and the
 solid lines, respectively.
The curves are shown for the cases $R_7/R_5=0.1, 0.3, 0.5, 0.7$ and $ 0.9$.
For each value of $R_7/R_5$, 
the inverse of 
 the radial norm is close to the corresponding transmission rate.
Therefore we can discuss the dependence of the transmission rate  on $R_7/R_5$ and $|E|/E_c$ by the radial norm.
For this purpose,
we show 
in Fig.\ref{rationorm}
 the radial norm
 for the negative incidence directions, $\Phi(kR_5,kr)$,  as a function of $r/R_5$ at six energies $E/E_c=kR_5/(2\pi)=0.1,0.3,0.5, 0.65, 0.8, 1.0$.
There are four regions of the parameter space which have
 different characters; they are  
region I $(1> r/R_5 >0.9; \,1 > |E|/E_c \geq 0)$ ,  
II $(0.9> r/R_5 >0.1; \,0.7> |E|/E_c \geq 0)$, 
 III $(0.9> r/R_5 >0.1;\, 1 > |E|/E_c > 0.7)$  
 and IV $(0.1> r/R_5 >0;\,1 > |E|/E_c \geq 0)$.
In the region I, 
all the curves in Fig.\ref{rationorm}
are very close to unity.
It indicates that the corresponding transmission rate  
is near unity and independent
 of $|E|/E_c$ as is shown by the lines of $R_7/R_5=0.9$
  in Fig.\ref{transnorm}.
In the region II, 
 the radial norm shows monotonic decrease
as either $r/R_5$ or $|E|/E_c$ increases.
The decrease with respect to $r/R_5$ 
becomes steep as $|E|/E_c$ decreases
and are almost proportional to $(R_5/r)^3$ when $|E|/E_c < 0.1$.
In the region III, 
however, the radial norm oscillates as $r/R_5$ increases.
When $|E|/E_c=1$ the radial norm has a minimum value around
 $r/R_5 = 0.5$.
Correspondingly, the curves
of $R_7/R_5=0.5$
 are larger than those of $R_7/R_5=0.3$ and 0.7 at $|E|/E_c=1$ in Fig.\ref{transnorm}.
It oscillates also with increasing $|E|/E_c$
when $0.4 > R_5 > 0.2$.
The lines of $R_5/R_7=0.3$ in Fig.\ref{transnorm}
shows the corresponding oscillations.
Around the boundary between the region II and III,
the decrease of the radial norm with increasing $r/R_5$ becomes critical
as is indicated by the plateau
of the curve corresponding to $|E|/E_c=0.65$ in Fig.\ref{rationorm}.
Correspondingly, all the curves for $R_7/R_5=0.7, 0.5$ and 0.3
 in Fig.\ref{transnorm} cross each other around the boundary energy
 $E/E_c=0.7$.
In the region IV,
the radial  norm is nearly proportional to $(R_5/r)^3$ 
 so that the transmission rate is
 almost  proportional to $(R_7/R_5)^3$ at each
 value of $|E|/E_c$.

The oscillating behavior in the region III can be interpreted
as a resonant effect in the following discussion.
The wave function propagates both in the radial and the angular
 direction.
Its wave length along the radial direction is assumed to be
 close to that of the plane wave obtained from the dispersion
 relation (\ref{dispersion}); the wave length is
 given by  $2\pi/k= R_5 |E_c/E|$.
When the length of the junction part along the radial direction,
i.e., $R_5-R_7$, coincides with the half of the wave length,
 the resonance occurs.
From this discussion, the condition of the resonance can be easily obtained
 as
\begin{equation}
R_7/R_5 = 1 -|E_c/2E| \;\;.
\label{resonance}
\end{equation}
The minimum points of the curves in Fig.\ref{rationorm} in the region III
correspond to this resonance following  approximately condition (\ref{resonance}); it comes around $r/R_5 = 0.5$
when $|E|/E_c=1$ and moves towards the left as $|E|/E_c$
decreases.
Around the resonance point, the transmission is almost perfect.
Such a resonance, however, does not occur in the region II and IV,
 since the radial norm becomes divergent when $kr$ approaches zero, i.e.,
when either $|E|/E_c=kR_5/(2\pi)$ or $r/R_5$ approaches zero.
This divergence comes from the terms of
 $kr N_2^2(kr)$ in eq.(\ref{norm}) which is almost proportional to $1/(kr)^3$
 for small $kr$ values.
Accordingly, 
 the transmission rate decays with the power law in
 proportion to $(R_7/R_5)^3$ in the region II and IV.

\section{Comparison between the analytical results and the numerical results}
Since the effective mass theory is an approximation of the tight
 binding model as is explained in the section III, its results have to
 be confirmed by comparing them to the corresponding numerical results
 from the tight binding model.
In this section, we shall compare the wave function calculated
 numerically by the tight binding model, $\vec{c}_j$,
 with those determined analytically  by eqs.(\ref{psiA}) and (\ref{psiB}),
 $\Psi$.
For the numerical calculation with the tight binding model, 
 the conditioned transfer matrix method
 explained in section II is used.
Figure \ref{ratiokp} shows the ratio
 of the corresponding quantities, $(|c^{(1)}_{j,i}|^2+ |c^{(2)}_{j,i}|^2)/(T\Psi)$  as a function of radial distance $r$, for the energy $E=-0.05|\gamma|$.
The values are plotted 
 for the (3,3)-(13,13) junction in Fig.\ref{ratiokp}-(a) and 
 for the (10,10)-(13,13) junction in Fig.\ref{ratiokp}-(b).
 Here  the incidence direction is  positive;
 $c^{(k)}_{j,i}$ is caused by the incident waves $\tilde{\psi}^R_{+k}$.
For the opposite incidence direction, similar results are also obtained.
Here the factor $1/T=|h_+|^2$ defined by eq.(\ref{hpm}) is necessary
  because the transmitted waves have unit flow in eqs.(\ref{psiA}) and 
 (\ref{psiB}), while  the incident waves have unit  flow
 in eq.(\ref{wavefunction+}).
One can see that the ratio is close to unit, i.e., 
 the coincidence is fairly good.
Correspondingly, the transmission rate by the conditioned transfer matrix
 and that by eq.(\ref{sigma}) coincide well with each other when
 $|E| < 0.2|\gamma|$, as shown in Fig.\ref{T13-13}.
The deviation from unity
becomes smaller as $|E|$ approaches zero, because all the
assumptions postulated in deriving eqs.(\ref{psiA}) and (\ref{psiB}) are
more valid for smaller $|E|$.
The deviations, however, become larger when $r \simeq R_5$ 
 in Fig.\ref{ratiokp}-(a),
while such a feature is not found in Fig.\ref{ratiokp}-(b).
The most natural explanation for this difference is
 that the deviations are caused  mainly by the evanescent waves in the tube
 parts,
 and their effects become more significant as
 their decay lengths along the tube axis become larger, i.e., as they
 approach the extended states.
The decay length  is determined by $l_{5,7}$ in eq.(\ref{tubebound})
 as $a/\sqrt{(2\pi  l_{5,7} a/R_{5,7})^2-\epsilon^2}$.
Thus the discussion below is concentrated to the most
 extended evanescent
 waves corresponding to
 $l_{5,7}=\pm1$.
When $R_5 \gg R_7$ as the case of Fig.\ref{ratiokp}-(a), the evanescent
 waves in the thicker tube side are much more extended and have 
larger effects than those in the thinner tube side.
As a result, the deviation near the thicker tube side
 is enhanced compared to that in the other place.
On the other hand, when $R_5 \simeq R_7$ as the case of Fig.\ref{ratiokp}-(b),
 the evanescent waves have similar decay lengths both 
 in the thicker and in the thinner tube side, so that
 the difference of the deviation does not depend on the sides.

To see the spatial variation  of the wave function,
 we show 
 $|\vec{c_j}|^2$ of the (3,3)-(13,13) junction
 calculated by the
 conditioned transfer matrix method 
 and the corresponding quantities
 determined analytically by eqs.(\ref{psiA}) and (\ref{psiB}).
The former is shown by  the closed symbols joined
 by the dotted lines and the latter is shown by
 the open symbols joined by the solid lines
  for $E= \pm0.2|\gamma|$ in Fig.\ref{normcj}-(a) 
 and  for $E= \pm0.23|\gamma|$ in  Fig.\ref{normcj}-(b).
The horizontal axis is the number of the bar $j$, 
 which are almost proportional
 to the radial distance $r$.
The diamonds and the squares correspond to the positive energy and the negative
 energy, respectively.
Fig.\ref{normcj}-(a) shows that the results by the two methods 
 agree quite well.
 Enhancement of the deviation at $j=16 \sim 20$ is caused by
 the evanescent waves.
In Fig.\ref{normcj}-(a), the wave function does not show monotonic decrease along the incidence
 direction but has a peak structure.
It gives rise to the resonant peak structures near $E=\pm0.2|\gamma|$ 
 in Fig.\ref{T13-13} which is already discussed in the previous section.
In contrast to the good agreement in Fig.\ref{normcj}-(a), 
the deviation becomes quite large in Fig.\ref{normcj}-(b)
 especially near $r = R_5$; 
the wave function
 by the conditioned transfer matrix grows more rapidly 
 than that by the analytical results with approaching the thicker tube side.
When $|E| \simeq E_c$, as in  the case of Fig.\ref{normcj}-(b), 
the decay lengths of the evanescent waves
 are very large, and cause the large discrepancies.
Corresponding to it,
large discrepancies occur also in the transmission rate; 
 only the data by the tight binding model show the sharp dips 
 near $E=\pm0.23\gamma$ in Fig.\ref{T13-13}.
The sharp dip appears approximately when $0.9 \leq |E/E_c| \leq 1$, i.e.,
 when the decay length of the evanescent waves in the thicker tube
 are larger than the diameter of the thicker tube.
Evanescent waves with their decay length
 larger than the diameter of the corresponding tube
 are called
 'quasi-extended' evanescent waves hereafter.

In Fig.\ref{ratiokp}, data corresponding to sites belonging to
 the pentagonal defect or the heptagonal defect are shown
by closed diamonds.
They indicate that the numerical 
 wave function calculated by the tight binding model
 is more localized at the defects than the analytical wave function.
Hereafter a localization strength for the $\theta$ direction
 at the defect is defined 
 as the ratio between the numerical norm per site at the defect and
 that of the bars the defect belongs to.
It is calculated by the tight binding model  and represented
 by
$(\sum_{i,k \in C_7} |c_{0,i}|^2 + |c_{1,k}|^2)(d_0+d_1)/(7|\vec{c}_0|^2+7|\vec{c}_1|^2)$
 at the heptagon
 and 
$(\sum_{i,k \in C_5} |c_{l_1,i}|^2 + |c_{l_1+1,k}|^2)(d_{l_1}+d_{l_1+1})
/(5|\vec{c}_{l_1}|^2+5|\vec{c}_{l_1+1}|^2)$
 at the pentagon with the notation of section II.
Here $i \in C_n$ means that site $i$ belongs to the $n$ membered ring defect.
Since the radial distance $r$ in the bar is almost constant,
the analytical quantity corresponding to the localization strength
has a maximum  represented
 by $\sum_{i=1}^2 (|X_{i,1}|+|X_{i,2}|)^2/ \sum_{i=1}^2(|X_{i,1}|^2
+|X_{i,2}|^2) \leq 2$  which is shown from eqs.(\ref{psiA}) and (\ref{psiB}).
Therefore we can say that the wave function is 'quasi-localized' at
the defect when the localization strength is larger than two.
The localization strength at the pentagon and that at the heptagon
 are shown as a function of the energy in Fig.\ref{pentagon}-(a) and
 Fig.\ref{pentagon}-(b), respectively.
 The solid lines and the dashed lines
 show the localization strength for the  the $(10,10)$-$(13,13)$ junction
 and  that for the $(3,3)$-$(13,13)$ junction, respectively.
 The solid lines shows that the the wave function is quasi-localized
 at both the defects
 when $|E| \simeq E_c$ and $R_7/R_5 \simeq 1$,
 i.e., when there are quasi-evanescent waves in both the tubes.
 Furthermore they also
 show asymmetry with respect to the energy axis;
 the quasi-localization at the pentagon (at the heptagon)  
 around $E = -E_c$ ($E = +E_c$ )  is larger
 than that around the energy with the opposite sign.
 These results indicate that the quasi-localization occurs due to the mixing
 between the quasi-extended evanescent waves
 and the 'defect levels'.
Here the 'defect levels' are defined as $n$ discrete energy levels,
 $-2|\gamma|\cos(2\pi\l/n)$  $(l=1,2,\cdots,n)$, of
an isolated $n$ membered ring calculated by the tight binding model. \cite{discl2}
Since the energy region around  zero is considered now,
the discrete level closest to $E=0$,
which is at $E \simeq -0.618|\gamma|$ for the pentagon,
and at $E \simeq +0.445|\gamma|$ for the heptagon, is the most important.
The closer distance in the energy
 between the two states enhances their mixing 
 so that it causes the asymmetry  with respect
 to the energy.
Since the defect levels are caused by the discreteness
 of the lattice,
 these results can not be reproduced by
 the effective mass equation.

 The energy where the quasi-localized evanescent waves
 in the thinner tube side appear is around $E = \pm E_cR_5/R_7$.
When $R_7/R_5 \ll 1$,
 it is distant from the considered energy region $|E|<E_c$
 so that there is no quasi-extended evanescent waves in the thinner tube side.
Though there are
 the quasi-extended evanescent waves in the thicker tube
 side,  the defect level of the heptagon
 cannot be mixed with them due to the large
 spatial distance between the heptagon and the thicker tube.
This is the reason why the dashed line in Fig.\ref{pentagon}-(b)
 do not show the quasi-localization.
In summary, we can say that the defect level is mixed
 with the quasi-extended evanescent waves
 only when they are close to each other both in the energy and in the space.

In this way, the analytical results by the effective mass approximation
 are not appropriate when the evanescent waves become quasi-extended.
The width of this energy region, however, is only about $0.1E_c$.
Except these narrow energy region, the analytical results are appropriate
 enough.

\section{Summary and conclusion}
In this paper, the junctions connecting the two metallic
nanotubes with different circumferences by a pair of
the pentagonal and the heptagonal defect are investigated.
Both  the wave function and the transmission
 rate are analytically obtained by the effective mass equation,
and a close relation between them is found.
To discuss the close relation,
the junction part is divided into pieces according to the radial 
 distance $r$ from the thinner tube side
 and the radial norm of the wave function
 is evaluated in each piece.
The radial norm is determined by the two parameters,
$|E|/E_c$ and $r/R_5$ where $E_c$ and $R_5$ are the threshold
energy and the circumference of the thicker tube, respectively.
The transmission rate approximates to the inverse
of the radial norm where $r$ is substituted with the circumference
of the thinner tube, $R_7$.
From the dependence of radial norm on the two parameters,
the parameter space is roughly classified into the four regions;
region I $(1> r/R_5 >0.9;\, 1 > |E|/E_c \geq 0)$ ,  
II $(0.9> r/R_5 >0.1;\, 0.7> |E|/E_c \geq 0)$, 
 III $(0.9> r/R_5 >0.1;\, 1 > |E|/E_c > 0.7)$  
 and IV $(0.1> r/R_5 >0;\,1 > |E|/E_c \geq 0)$.
 In the region I, the radial norm is close to unity
 so that the corresponding transmission is almost
perfect, independent of $|E|/E_c$.
In the region III, the radial norm oscillates near unity
as the function of either of the two parameters.
The period of the oscillation with respect to $r$ 
is approximately the same as half wave
length.
Here the wave length is that
 of the plane wave, i.e., 
 it is obtained  from the liner dispersion relation
 of the monolayer graphite, eq.(\ref{dispersion}).
 Almost perfect transmission due to the resonance occurs
 in region III
when the radial length
 of the junction part, $R_5-R_7$, coincides with 
 this period, because the corresponding radial
 norm has a minimum value there.
As either $|E|/E_c=kR_5/(2\pi)$ or $r/R_5$ decreases,
however, the radial norm becomes divergent owing to
 the term $kr N_2^2(kr)$ so that the corresponding
 transmission rate approaches zero.
It leads to the power law decay of the transmission
rate proportional to $(R_7/R_5)^3$  in the region II
and IV instead of the resonant feature.\cite{tamurajunction1,tamurajunction2,pre}

 The nanotube junction can be considered
 as a combined system, that is 
 the quasi two-dimensional structure (= the junction part)
 connecting the two quasi one-dimensional structures (= the tube parts).
The wave functions in the former part become the 2D waves,
 whose radial parts are given by the sum of $J_i(kr)$ and $N_i(kr)$ $(i=1,2)$, 
 while they are the 1D plane waves in the tube parts.
Nevertheless, the condition of the resonance 
 in the region III, eq.(\ref{resonance}),  is obtained
from the wave length of the 1D plane wave.
It indicates that the four components of the wave function in the junction
 part, $J_i(kr)$ and $N_i(kr)$ $(i=1,2)$, 
 are combined appropriately
 to be fitted well with the 1D plane waves in the tube parts.
On the other hand, the wave function shows the power law decay
along the radial direction in the region II and IV,
 because only the component $N_2(kr)$ becomes dominant.
In other words, the wave function in the junction part has the 
two dimensional character in the region II and IV 
so that its matching with 
the plane wave in the tube parts becomes worse
 than in the region III.
The wave function in the junction part has different dimensionalities
 in this way.

The wave functions analytically obtained by the effective mass equation
are compared with that numerically obtained by the tight binding model.
When $|E|$ is very close to $E_c$, the discrepancies between them
becomes considerable.
From the dependence of the discrepancies on the radial distance,
we speculate that they are mainly due to the evanescent
 waves with their decay length larger than
 the diameter of the corresponding tube
 and they are called quasi-extended evanescent waves.
The quasi-extended evanescent waves
 cause the sharp dip of the transmission rate,
which is absent in the analytical result.
When the quasi-extended evanescent waves and the energy levels
 of the $n$ membered ring defect $(n=5,7)$
 are close both in the energy and in the space,
mixing between them appears as the enhancement of
the wave function at the corresponding defect.
Nevertheless coincidence between the analytical wave functions
 and the numerical ones are fairly good except
 the narrow energy region $0.9 < |E/E_c| < 1$.

The nature of the wave function discussed in this paper
 can be observed
 by Scanning Tunneling Microscopy (STM), because
 STM images 
 reflect  the Local Density of States (LDOS),
 which are proportional to the sum of the squared 
 wave functions,  eqs. (\ref{psiA}) and (\ref{psiB}),
 in the nanotube junction part.
 The third terms in eqs. (\ref{psiA}) and (\ref{psiB})
 give rise to the $\sqrt{3} \times \sqrt{3}$ pattern
 or the oscillation along the $\theta$ direction
 in the STM images.
 They fade away as $|E|$ approaches zero.
It is also expected that the images depend on 
 the direction of the electronic current 'along the tube axes',
 which should not be confused with that of the tunneling current
 'from the STM tip to the sample'.
When $R_5/R_7 \gg 1$ and $E \simeq 0$, for example, 
 the squared absolute value of the wave function
 in the junction part
 is almost proportional to $r^2$  when  the current flows
 from the thicker tube to the thinner tube
  and  to $1/r^4$ when it flows along the inverse direction.
We expect that these results will promote the investigation
 of the nanostructures composed by the nanotubes including
 the nanotube junctions.

\section*{Acknowledgment}
This work is supported in part by the Core Research for Evolutional 
 Science and Technology (CREST) of the Japan Science and Technology 
 Corporation(JST).

\section*{Appendix}
In the tight binding model, the probability flow
  from site $i$ to site $j$ is
 represented by ${\rm Im}(a_i^*H_{i,j}a_j)$, where $a_i$ is
 the amplitude of the wave function at site $i$ and $H_{i,j}$
 is the hopping integral connecting site $i$ and $j$.
 Conservation of the flow is guaranteed by the tight binding
 equation, $E a_i = \sum_j H_{i,j} a_j$.
 In this paper, $H_{i,j}$ is nonzero
 only when $i$ and $j$ are nearest neighbors.
 Consider bars for $j \leq 0$ forming to the $(m,n)$ nanotube
 part in Fig.\ref{tenkaibars}  .
 The amplitudes of wave functions are represented by $r_{j,i}$
 and $l_{j,i}$ in the same
 way as Fig.\ref{rlbars}.
 Then the flow between $j-1$'th bar and $j$'th bar
 is represented by $\gamma \sum_{i=1}^{m+n} {\rm Im}(r^*_{j-1,i}l_{j,i})$.
 Hereafter the common hopping integral between the nearest
 neighbors $\gamma$ and the lattice constant $a$ is chosen to be units.
The flow $J$ corresponding to $\vec{\psi}_{K+}$ in (\ref{propK})
is $J = m\sin ( \eta-k_2+2\pi/3) + n\sin ( -\eta+k_1+2\pi/3)$,
 where the first term comes from the $m$ bonds, $1 \leq i \leq m$,
 and the second term comes from the other $n$ bonds.
Here $k_i= \vec{k}\cdot \vec{e}_i$ and $\eta$ is the angle of $\vec{R_7}$
 with respect to $x$ axis, i.e., $(\cos (\eta), \sin(\eta))= (\sqrt{3}(m+n)/(2R_7), (n-m)/(2R_7))$ and $R_7=\sqrt{m^2+n^2+mn}$.
By developing about $k_i$ to the first order and 
 using the boundary condition $mk_1+nk_2=0$,
the flow $J$ is represented as
\begin{equation}
J = R_7 - \frac{\sqrt{3}k_1}{2R_7}m(n+m) +O(k_1^2) + O(k_2^2)\;\;.
\label{flowr7}
\end{equation}
The absolute value of the second term in (\ref{flowr7})
is less than $\sqrt{3}\pi R_7/R_5$, since the energy region where
 the  channel number is kept to two, i.e., $k R_5 \leq 2\pi$,
 is considered now.
It can be deduced from it that
  the flow is  the almost constant value, $R_7$,  when
 $R_5 \gg \sqrt{3}\pi$.
On the other hands, the value of $|\vec{\alpha}_{7+}|^2$ 
obtained by substituting $F$'s in eq.(\ref{alpha})
 with those of eq.(\ref{propK}) for $\vec{\psi}_{K+}$  is also $R_7$.
Therefore the normalization factor in eq.(\ref{alpha}) is correct.
The normalization factor of Ref.\cite{pre} and Ref.\cite{pre2}
 is different from it  by a factor $\sqrt{2}$.
This difference does not change the transmission rate,
 but  when ratio between the wave function in the tube parts and that in
 the junction part is considered,
the normalization factor in eq.(\ref{alpha}) have to be used.

\begin{figure}
\caption{
 The development map showing the bond network of the (2,2)-(2,5) junction.
 The filled circle at the bottom and the open circle at the top in each bar
 indicate an identical atom.
 To form the junction, the development map is rolled up so that
 the filled circle and the open circle coincide with each other 
 in each bar.
 The solid lines and the dashed lines represent the bonds within
 the bar and those connecting the neighboring bars, respectively.
The points $P_5(=Q_5)$, $P_7(=Q_7)$ and $O$ are the centers of the
pentagonal defect, that of the heptagonal defects and the origin of
 the coordinate, respectively, which are shown also
in Fig. \protect\ref{junctiontenkai}.
The two sets of the translation vectors, $\{ \vec{e}_1,\vec{e}_2\}$
and  $\{ \vec{e'}_1,\vec{e'}_2\}$, and corresponding sets of
orthogonal vectors  $\{ \vec{e}_x,\vec{e}_y\}$ and  $\{ \vec{e}_{x'},
\vec{e}_{y'} \}$ are shown.
Note that  definition of sublattices $A$ and $B$ are changed between the two sets.}
 \label{tenkaibars}
\end{figure}

\begin{figure}
\caption{Notation to represent the amplitudes of the wave function.
 It is illustrated with the fist bar and the second bar 
 in Fig.\protect\ref{tenkaibars}.
The site in the $j$'th bar can be classified to the two
 groups according to whether it connects
 with the right $(j+1)$'th bar
 or with the left $(j-1)$'th bar.
 The former and the latter are denoted by $r_{j,i}$
 (closed triangles) and $l_{j,i}$ (closed squares),
 respectively, when the corresponding site has the bond $i$, which
 connects the neighboring bars and is numbered from the bottom to the top.}
\label{rlbars}
\end{figure}

\begin{figure}
\caption{Development map of the nanotube junctions.
The lines '${\rm E_7P_7P_5E_5}$' 
 are connected and become identical with the lines 
'${\rm F_7Q_7Q_5F_5}$', respectively.
The rectangles '${\rm E_7P_7Q_7F_7}$' and '${\rm P_5E_5F_5Q_5}$' form the thinner tube and the thicker tube, respectively. 
The triangle '${\rm OP_7P_5}$' is the same as the '${\rm OQ_7Q_5}$'
 rotated by   60 degrees. The quadrilateral '${\rm P_7P_5Q_5Q_7}$' forms 
 a junction part with a shape of a part of a cone. 
A heptagonal defect and a pentagonal defect are introduced
at ${\rm P_7(=Q_7)}$ and ${\rm P_5(=Q_5)}$, respectively.
The direction of the circumferences of the tubes in the development map
 is represented by their angles $\eta_5$ and $\eta_7$ measured anti-clockwise
 with respect to the $x$ axis which is defined
 in Fig. \protect\ref{tenkaibars}.
The angle between the axes of the two tubes, $\phi$, is defined as
 $\phi=\eta_7-\eta_5$.}
\label{junctiontenkai}
\end{figure}

\begin{figure}
\caption{Transmission rates per channel, $T$,  as a function of the energy.
The transmission rate calculated by eq.(\protect\ref{sigma}) 
are shown
by the solid lines, and
 those calculated 
 by the tight binding model 
 are shown
 by the dotted line for the (10,10)-(13,13)
 junction and by the dashed line for the (3,3)-(13,13) junction.
The considered energy region is  $|E|< E_c$  where
 the channel number is kept to two in 
 both the tube parts.
The threshold energy $E_c$ for the effective mass theory is
  about $0.242|\gamma|$ and that for the tight binding model is
 about $0.239|\gamma|$,
 where $\gamma < 0$ is  the hopping integral.
One can easily see the agreement between the two methods
 is fairly good.}

\label{T13-13}
\end{figure}

\begin{figure}
\caption{
Schematic development maps representing the relation
 between the radial norm $\Phi$ and the transmission rate $T$.
 The thicker tube and the region CC'D'D are common in both the development
 maps}
\label{scheme}
\end{figure}
\begin{figure}
\caption{Inverse of the radial norm for the positive incidence direction
defined by $2/(\Phi(kR_7,kR_5) +1)$ , and that for the negative incidence direction $2/(\Phi(kR_5,kR_7)+1)$ are shown by the dashed lines and the dotted lines, respectively.
Here the radial norm $\Phi$ is defined in eq.(\protect\ref{norm}).
Values of $R_7/R_5$ are 0.1, 0.3, 0.5, 0.7, and 0.9,
 which are attached to the corresponding lines.
The  transmission rate per channel calculated by eq.(\protect\ref{sigma}) 
is shown by the solid lines.
 It can be seen that the radial norm gives a good estimated value
 of the transmission rate.
 The horizontal axis is 
the energy normalized by $E_c$. Here $|E|< E_c$ represents
 the energy region where the channel number is kept to two in 
 both the tube parts.}
\label{transnorm}
\end{figure}

\begin{figure}
\caption{
Radial norm for the negative incidence direction,
 i.e., $\Phi(kR_5,kr)$  is shown
as a function of $r/R_5$ for
 five energies, $kR_5/(2\pi)=E/E_c=0.1, 0.3, 0.5,0.65, 0.8$ and 1.0.
Here $\Phi$ is defined by eq.(\protect\ref{norm}).
The values of $E/E_c$ are attached to the corresponding lines.
 Both the axes are represented with log scale.
 Inset shows the regions of the parameter space, I, II, III and IV,
 which are used in the discussion in the text.}
\label{rationorm}
\end{figure}

\begin{figure}
\caption{Comparison
 between the wave functions  $|c^{(k)}_{j,i}|^2$ 
calculated by the tight binding model,              ,  
and $\Psi$ defined by eqs.(\protect\ref{psiA}) and (\protect\ref{psiB});
It shows $(|c^{(1)}_{j,i}|^2+ |c^{(2)}_{j,i}|^2)/(T\Psi)$ 
 as a function of radial distance, $r$, (a) for the (3,3)-(13,13) junction
 and (b) for the (10,10)-(13,13) junction.
 They are shown for $j=1,2, \cdots, l_1$.
The energy is $E= -0.05|\gamma|$.
 Here  $|c^{k}_{j,i}|^2$ 
 is caused by the incident waves with positive direction, 
$\tilde{\psi}^R_{k+}$.
Here $1/T=|h_+|^2$ defined by eq.(\protect\ref{hpm}) is used
 for the normalization.
The vertical axis is represented with log scale.
Data at the pentagon and at the heptagon are shown by the closed diamonds. }
\label{ratiokp}
\end{figure}

\begin{figure}
\caption{ Norm
 of the wave function of each bar, $|\vec{c_j}|^2$, 
 of the (3,3)-(13,13) junction
(a) for $E= \pm0.2|\gamma|$ and (b) for $E= \pm0.23|\gamma|$.
The horizontal axis represents  the position of the bar $j$ 
 shown in Fig.1.
The incidence direction is positive, i.e., from the right
 to the left in this figure.
The dotted lines with closed symbols and the solid lines with open
 symbols correspond to those calculated by eq.(\protect\ref{wavefunction+})  
and those calculated by eqs.(\protect\ref{psiA}) and (\protect\ref{psiB}), 
respectively.
The diamonds and the squares correspond to the positive energy and the negative
 energy, respectively.}
\label{normcj}
\end{figure}

\begin{figure}
\caption{
The localization strength for the $\theta$ direction
 (a) at the pentagon and (b) at the heptagon
 as a function of the energy.
The incidence direction is positive.
 The solid lines and the dashed lines
 correspond to  the $(10,10)$-$(13,13)$ junction
 and  the $(3,3)$-$(13,13)$ junction, respectively.}
\label{pentagon}
\end{figure}


\begin{references}
\bibitem{noteepsilon}
Though the sign $\epsilon/|\epsilon|$ is absent in Ref.\cite{pre},
 it is necessary.
This difference does not change the transmission rate,
 but it must be considered when the wave function is
 calculated.
\bibitem{note}
When the probability flow of the wave function
 $\sum (x_i \psi_i + y_i \psi_i^*)$ becomes
 $\sum (|x_i|^2 - _|y_i|^2) v_i$ where
 cross terms like $x^*y$ and $y^*x$ do not appear,
 we call $\{ \psi_i \}$ is "orthogonalized".
Here "orthogonalized" does not mean  that  for usual inner products
 of the wave functions $\int dV \psi_i^*(x)\psi_j(x)$,
 but means that for the probability flow, $\int dS \hat{J}(\psi_i, \psi_j)$,
 where $\hat{J}$ is the operator of the probability flow defined in Appendix
 and $\int dS$
 means the integration about the cross section of the system.
Furthermore $\psi$'s are normalized so that  $v_i$ becomes  a constant
 independent of $i$.
Such a set of "orthonormalized" wavefunctions $\{ \psi_i \}$, generally speaking, are not orthonormal in a usual meaning, i.e.,
 $\int dV \psi_i^*(x)\psi_j(x)  \neq \delta_{ij}$.
\bibitem{note2}
Note that in general problems, orthogonality of the incident 
 waves and that of the transmitted waves
 hold in the same time only when
 the incident waves are chosen to be 'eigen channels'
 discussed in Ref.\cite{brandbyge}.
\end{references}
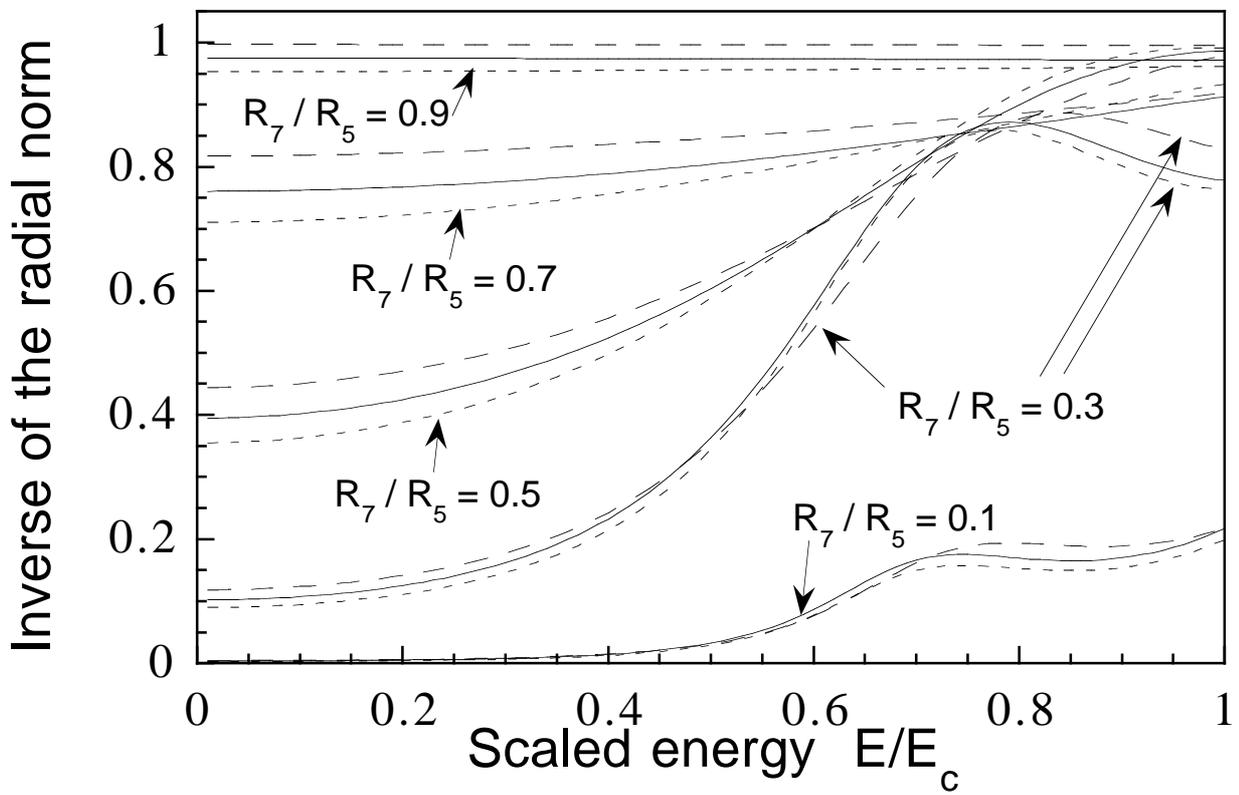
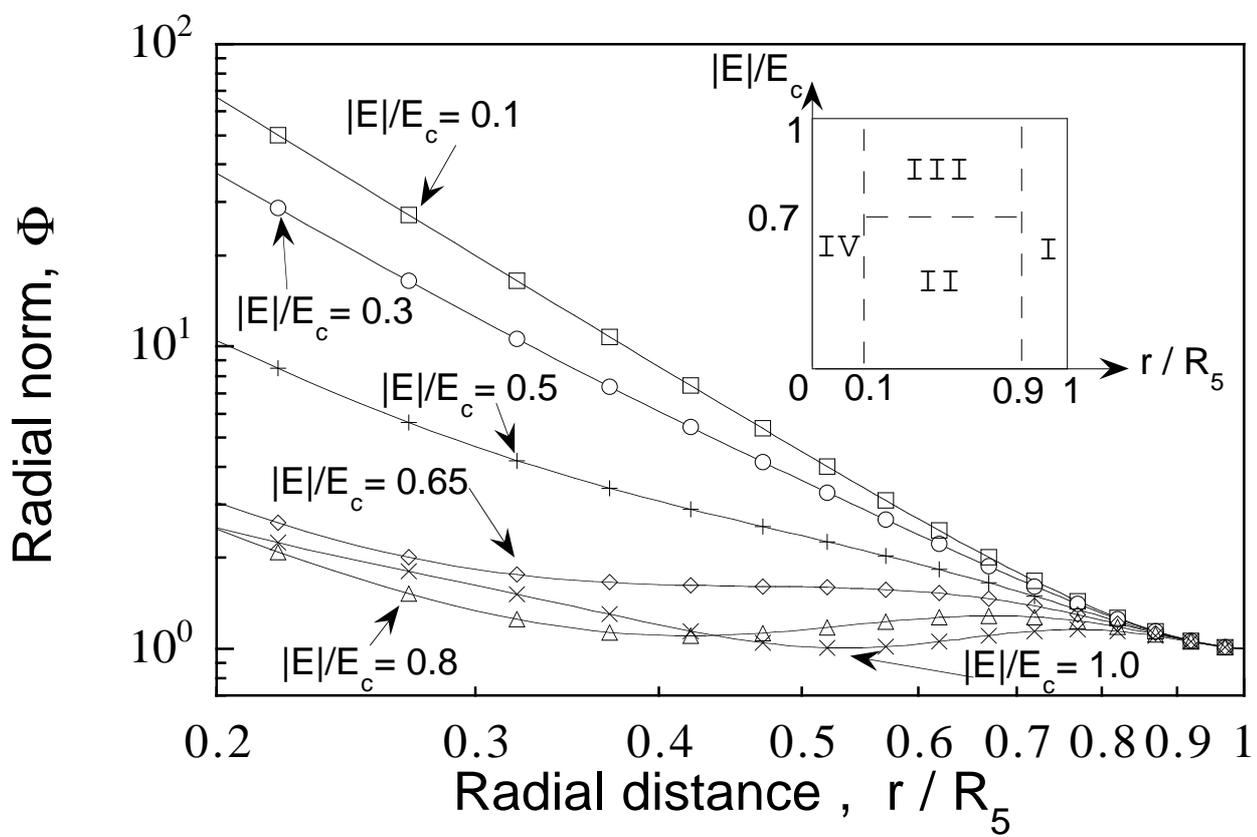
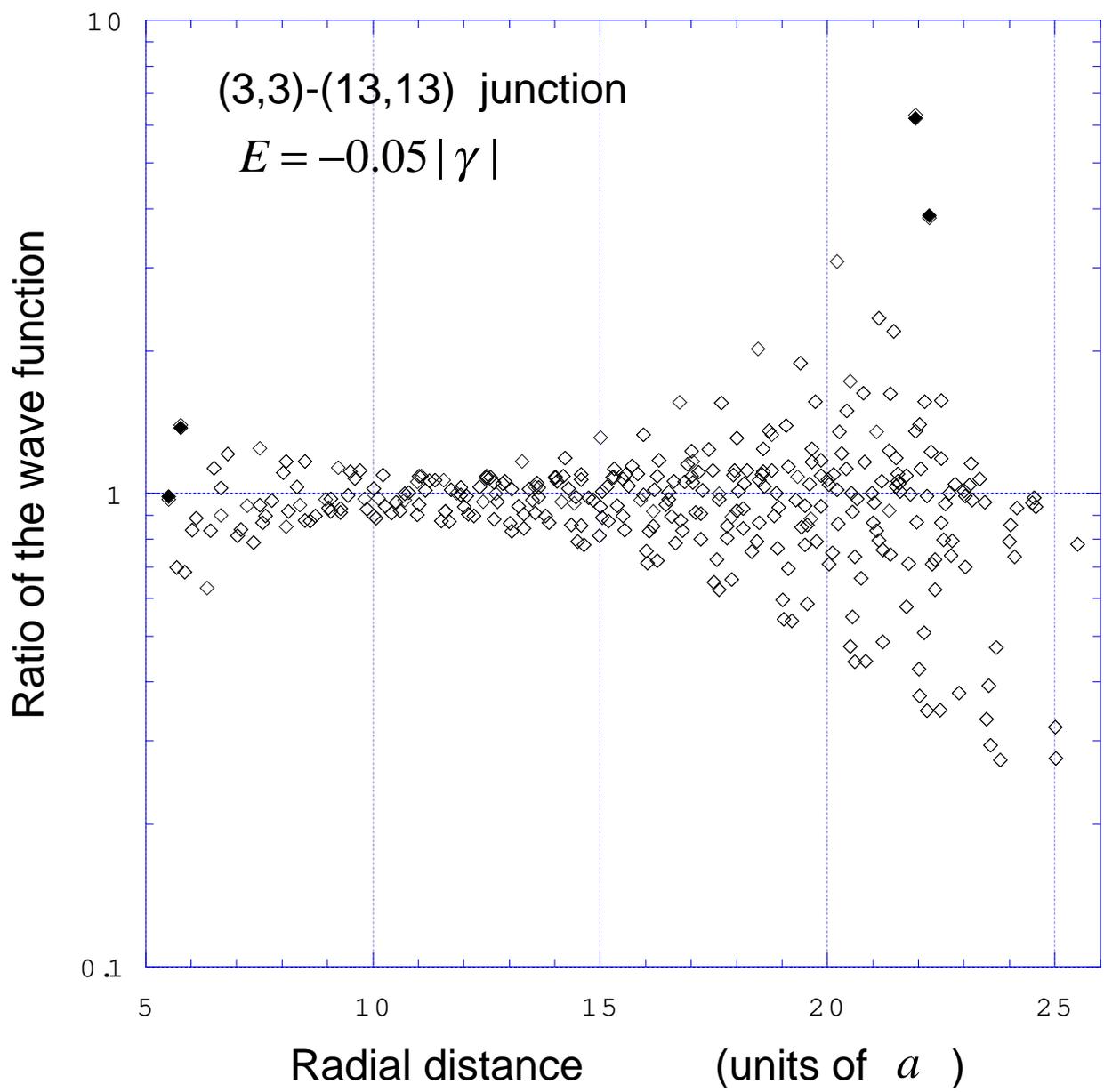
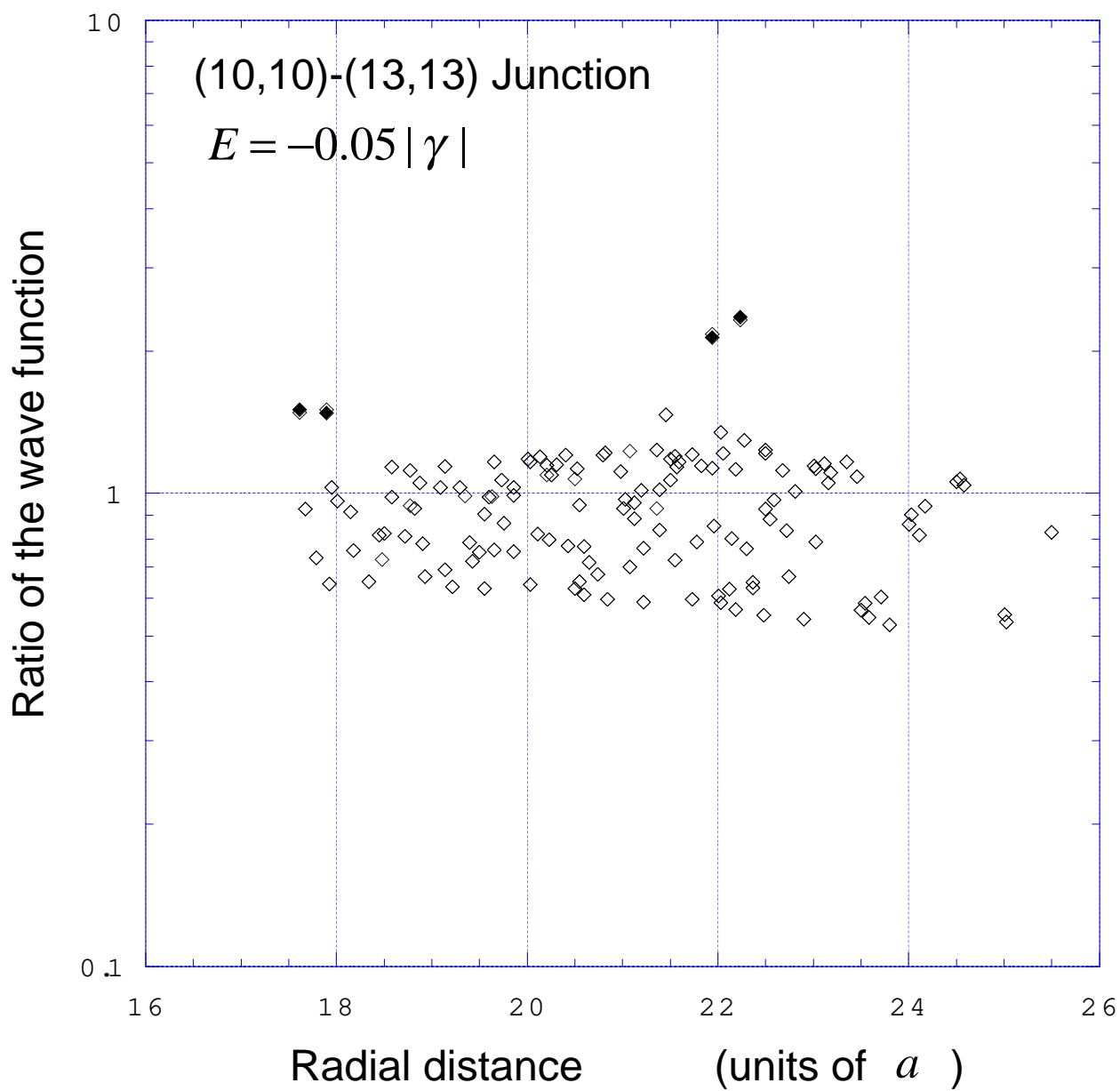
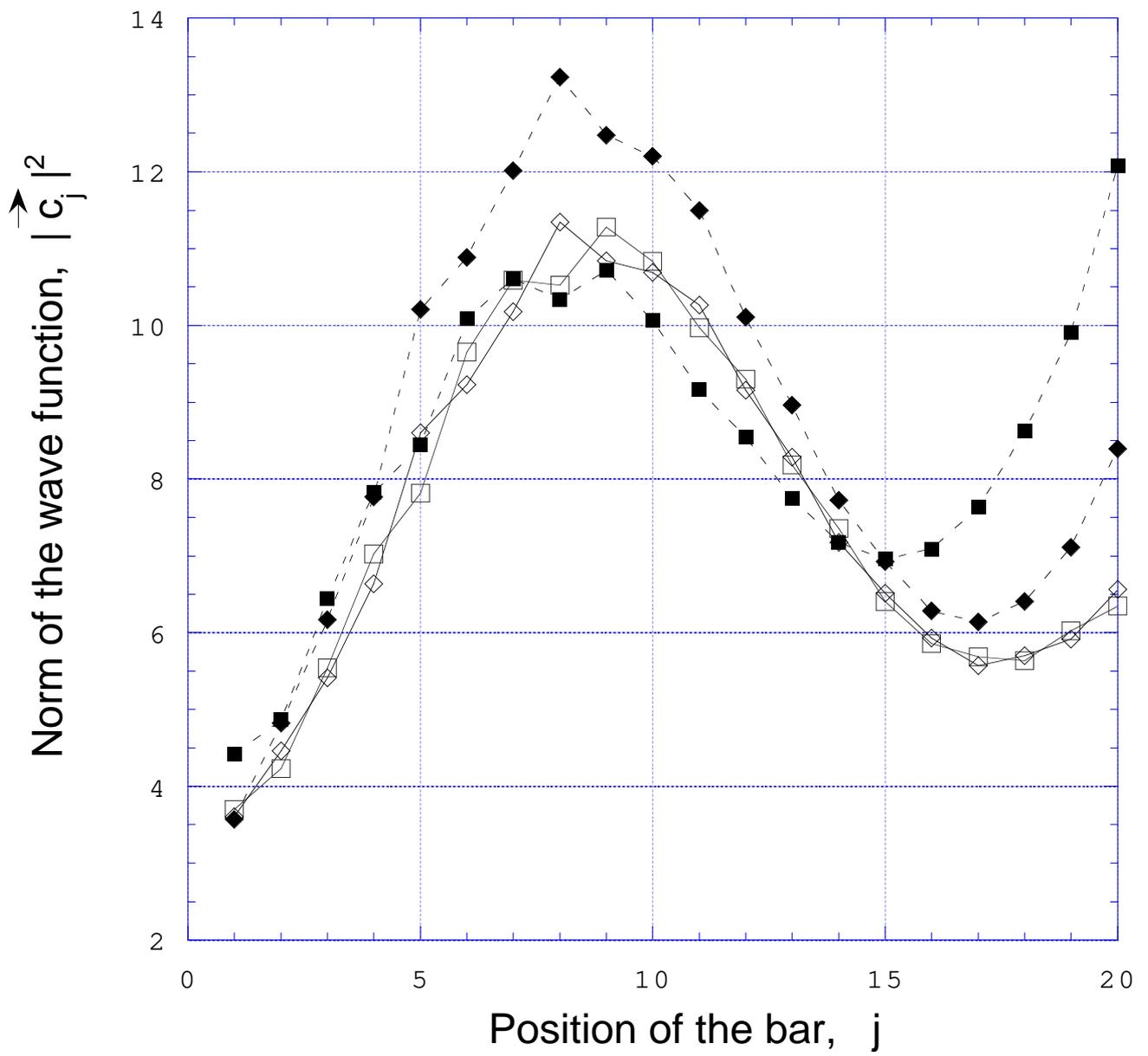
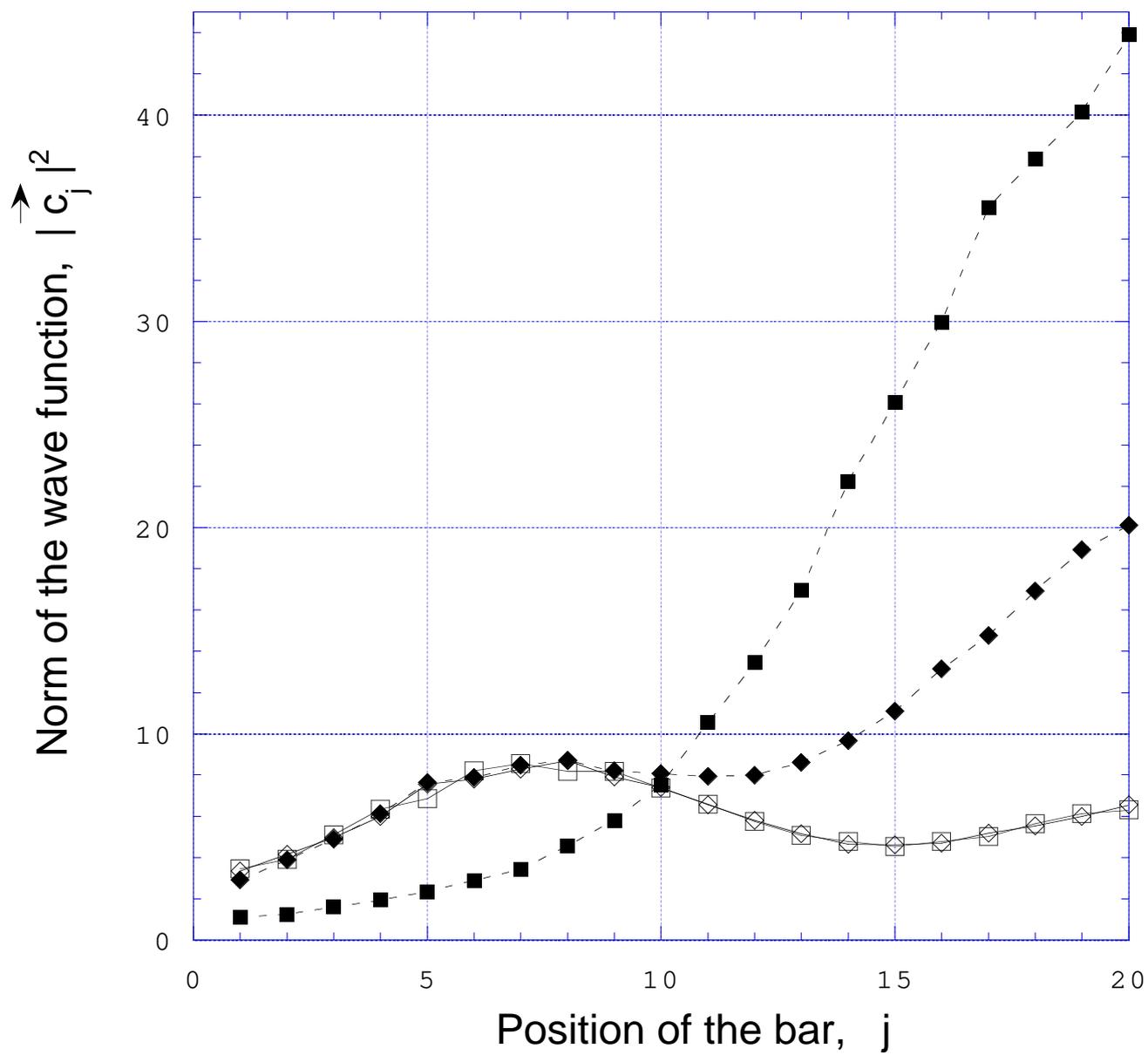
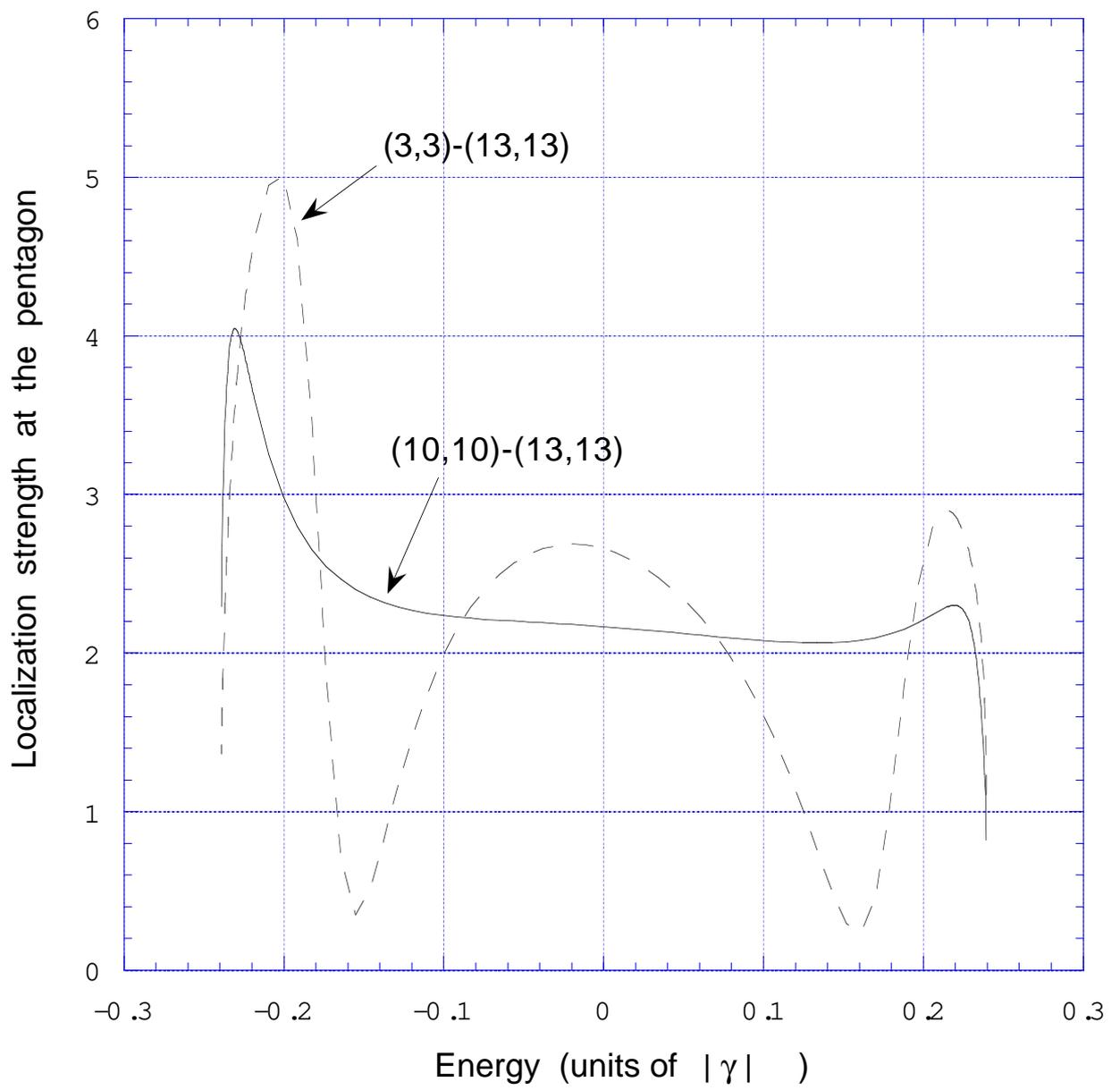
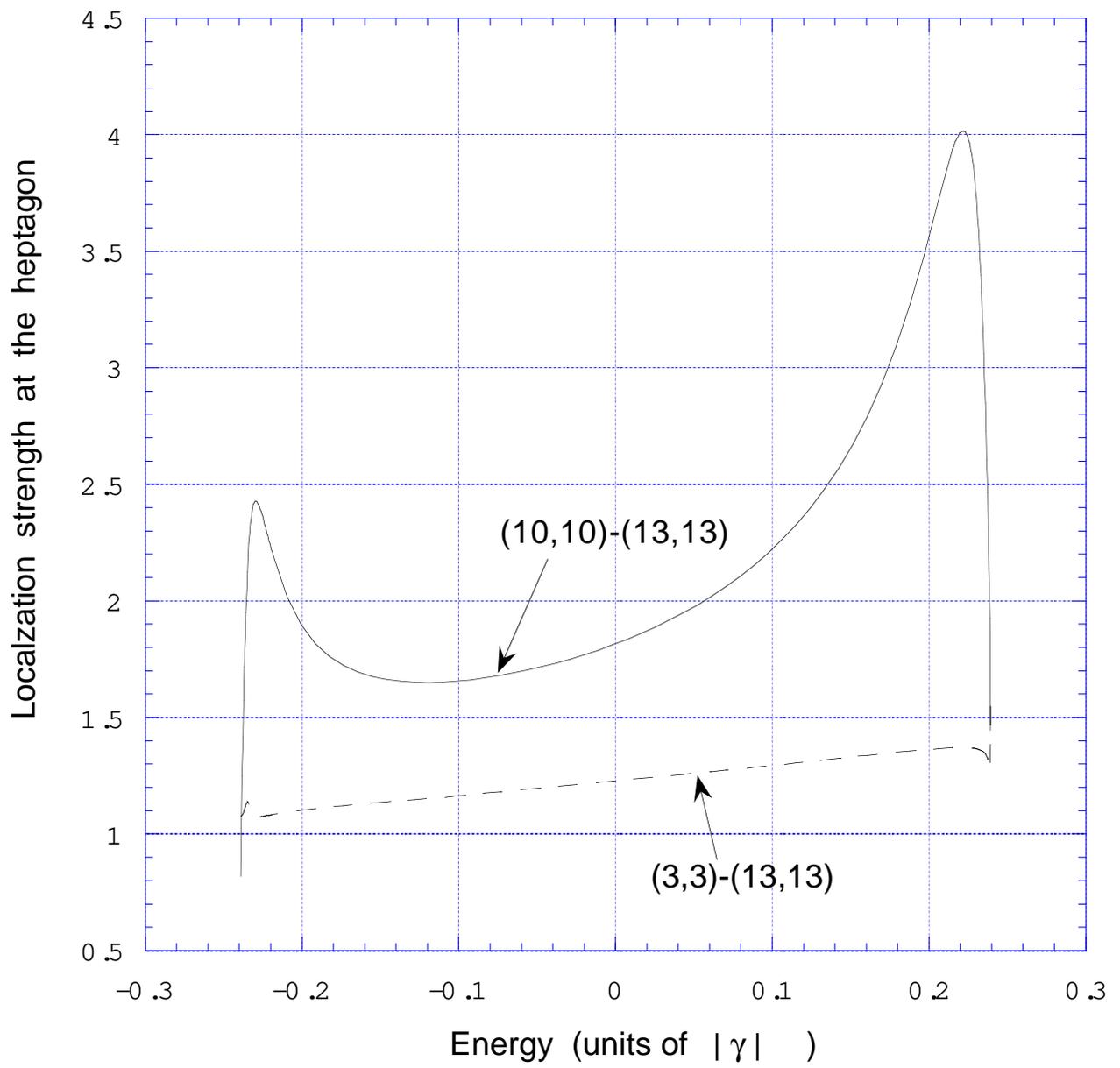
\end{document}